%% file: arXiv.tex
\documentclass[aps,jcp,superscriptaddress,amsmath,amssymb]{revtex4-2}
\setcitestyle{super}

\usepackage{dcolumn}
\usepackage{bm}

\newcolumntype{.}{D{.}{.}{8}}
\usepackage[english]{babel}
\usepackage[utf8]{inputenc}
\usepackage{physics}
\usepackage{braket}
\usepackage{tabularx,booktabs,array,dcolumn}
\usepackage{setspace}
\usepackage{xcolor}
\usepackage{graphicx,psfrag,subfigure}

\usepackage{float}
\usepackage[all]{xy}
\usepackage{multirow}

\usepackage{amsmath}
\usepackage{amstext}
\usepackage{amsfonts}
\usepackage{amssymb}
\usepackage{color}
\usepackage{longtable}
\usepackage{rotating}
\usepackage{multirow}
\usepackage{tikz}
\usepackage[version=4]{mhchem}
\usepackage{listings}
\usepackage{float}
\usepackage{threeparttable}
\usepackage{url}
\usepackage{here}
\usepackage[absolute,overlay]{textpos}
\usepackage{siunitx}
\usepackage{mathrsfs}
\usepackage[inline]{enumitem}
\usepackage{mhchem}

\usepackage[unicode]{hyperref}
\hypersetup{
   unicode=true,          
   plainpages=false,
   colorlinks=true,       
   linkcolor=blue,          
   linkcolor=blue,          
   citecolor=blue,        
   filecolor=blue,      
   urlcolor=blue           
}
\newcommand{\be}{\begin{equation}}
\newcommand{\ee}{\end{equation}}
\newcommand{\bea}{\begin{eqnarray}}
\newcommand{\eea}{\end{eqnarray}}
\newcommand{\Eq}[1]{Eq.~\ref{#1}} 
\newcommand{\FIG}[1]{Fig.~\ref{#1}} %

\newcommand{\bos}[1]{\boldsymbol{#1}}
\newcommand{\mr}[1]{\mathrm{#1}}
\newcommand{\cm}{$\text{cm}^{-1}$}

\newcommand{\Cs}{$C_\mathrm{s}$}
\newcommand{\Ctv}{$C_{\mathrm{3v}}\mathrm{(M)}$}
\newcommand{\Vp}{O_\mr{pr}}

\newcommand{\SM}{Supplementary Material}

\newcommand{\pol}{$\alpha_{ij}$}

\begin{document}

\title{%
Vibrational infrared and Raman spectra of the methanol molecule with equivariant
neural-network property surfaces 
}

\author{Ayaki Sunaga}
\affiliation{%
ELTE, E\"otv\"os Lor\'and University, Institute of Chemistry, P\'azm\'any P\'eter s\'et\'any 1/A 1117 Budapest, Hungary
}
\author{Albert P. Bartók}
\affiliation{%
Department of Physics, University of Warwick, Coventry, CV4 7AL, UK
}
\affiliation{%
Warwick Centre for Predictive Modelling, School of Engineering, University of Warwick, Coventry, CV4 7AL, UK
}

\author{Edit M\'atyus}%
\email{edit.matyus@ttk.elte.hu}
\affiliation{%
ELTE, E\"otv\"os Lor\'and University, Institute of Chemistry, P\'azm\'any P\'eter s\'et\'any 1/A 1117 Budapest, Hungary
}%

\date{30 April 2026}

\begin{abstract}
\noindent 
Electric dipole and polarizability surfaces are developed for the methanol (\ce{CH3OH}) molecule using 
\textit{ab initio} electronic structure data, computed at the CCSD/aug-cc-pVTZ level of theory, and equivariant neural networks. 
These property surfaces are used to compute vibrational infrared and Raman intensities with variational vibrational energies and wave functions. 
The energies and wave functions, fully accounting for the large-amplitude motion and tunneling splitting states, are from continued variational vibrational computations, based on earlier work [Sunaga \textit{et al., J.~Chem. Phys.}, 2025, \textbf{163}, 064101], up to 3700~cm$^{-1}$\ beyond the zero-point vibration, now reaching the O-H stretching fundamental. All vibrational fundamentals, combination and overtone bands are in excellent agreement with available (gas-phase) experimental data, with a 2.2~cm$^{-1}$\ root-mean-squared deviation of the fundamentals from experiment. 
These developments constitute an important step towards a quantitative and comprehensive exact quantum dynamics model of the methanol molecule, and a linelist for astrophysical applications.
\end{abstract}

\maketitle 


\clearpage
%
%
\section{Introduction}
\noindent %
Alongside the (ro-)vibrational energy intervals, the peak intensity is an important element for understanding molecular spectra and dynamics, and for using this information to trace the chemical composition of outer space.\cite{Exomol1,Exomol2} 
To compute transition intensities, here we consider infrared (IR) and Raman intensities, we need high-quality representations of not only the potential-energy, but also the electric dipole and polarizability surfaces. 

It has long been recognized that the permutational invariance of atomic nuclei is an essential feature of a mathematical representation of the potential energy surface (PES),\cite{Brown2003JCP_CH5+,Huang2005JCP,
Manzhos2006JPCA_identical_nuc_NN,Behler2007PRL_identical_nuc_NN,
Braams2009IRPC_pip,Jiang2013JCP_NN-PIP,Qu2018ARPC_PIP,Bowman2025JCP_PIP_review} and also the dipole moment surface (DMS).\cite{Huang2005JCP,Wang2009JCP_DMS} 
Traditionally, the polynomial expansion has been employed for PESs~\cite{Brown2003JCP_CH5+,Huang2005JCP,Braams2009IRPC_pip,Qu2018ARPC_PIP,Bowman2025JCP_PIP_review} and DMSs.\cite{Jensen1988JMS_CH2,Huang2005JCP,Wang2009JCP_DMS,Lodi2011JCP_water,Huang2025JQSRT}
Recent advances in equivariant representation techniques\cite{Bartok2010arXiv_rot_invariant,Batzner2022NC_NequIP,Batatia2022_proc_MACE,DoBiPeCe25} enable the construction of functions that are both permutationally invariant and correctly preserve spatial rotation-inversion properties for tensors. 
Recently, machine-learning-based fitting techniques such as high-dimensional neural network potentials,\cite{Behler2011JCP_HDNNP,Behler2021CR_HDNNP} Gaussian approximation potentials,\cite{Bartok2010PRL_GAP,Bartok2010arXiv_rot_invariant} atomic cluster expansion,\cite{Drautz2019PRB_ACE} graph neural network~\cite{Batatia2022_proc_MACE} have been developed.

Methanol is one of the smallest prototypes for a polyatomic molecule with one large-amplitude motion. The coupling of small-amplitude vibration, torsion, and rotation has been long studied by high-resolution spectroscopy.\cite{Serrallach1974JMS,Hanninen1999JCP_fit,Moruzzi1995_ch3oh,Xu1997JMS_CH_strech,Hunt1998JMS_OH_stretch,Wang1998JCP_fit,Lees2002PRA,Temsamani2003JMS,Lees2004JMS,Harrison2012JQSRT_methanol_cross_section,Konnov2025JQSRT_methanol} 
Methanol has been proposed as a sensitive probe, exploiting internal rotation, to detect variations in the proton-to-electron mass ratio.\cite{Jansen2011PRL_ch3oh,Levshakov2011APJ_methanol,Bagdonaite2013Science,Vorotyntseva2024MNRAS_ch3oh}
Recently, precision spectroscopy studies of methanol have been conducted using frequency combs.\cite{Santagata2019optica,Tran2024APL}
In parallel, a theoretical proposal for detecting parity-violation in molecules targeted halogen-substituted methanol molecules and identified the most promising candidate from this family.\cite{sunaga2025JCP} 
The transition intensities of methanol (and isotopologues) have been proposed to probe the physical conditions in outer space, such as temperature.\cite{Muller2021AA_ch3oh_temperature,Evans2025AstrophysJ_ch3oh_temperature}

Besides optical and astrophysical uses, a comprehensive vibrational dataset (including transition properties computed in this work) may help us understand positron annihilation spectra of methanol.\cite{Gribakin2017PRA_methanol} Mode combination and overtone vibrations---usually faint in regular IR spectra---have been recently proposed as key contributors to positron annihilation spectra via vibrational Feshbach resonances.\cite{Gribakin2010RMP,Gilbert2002PRL_pos}

Developments in exact quantum dynamics have focused on polyatomic systems with large-amplitude motions. Recent work extended molecular complexity up to malonaldehyde,\cite{Lauvergnat2023CPC_malon} handled non-Abelian symmetry, such as in acetonitrile,\cite{Rey2024JCP_nest_tensor} and examined intra- and intermolecular quantum dynamics in molecular complexes.\cite{Viglaska2025JCP_contraction,Simko2025JCP_H2O_tri,Felker2025JCP_h2o_trimer,Jing2025JPCL_Break_1cm} Further methodological advances elaborated tensor-train\cite{Rey2024JCP_nest_tensor}, monomer-based contractions,\cite{Viglaska2025JCP_contraction,Simko2025JCP_H2O_tri,Felker2025JCP_h2o_trimer} truncated basis and grid representations,\cite{Lauvergnat2023CPC_malon} novel collocation techniques,\cite{wodraszka2024JCP_collocation} and extension of the $n$-mode representation beyond normal coordinates.\cite{BaLaCh24}  

This work is the third piece in a recent series on the methanol molecule; first, the coordinate definition and \ pruned variational vibrational computations were elaborated;\cite{Sunaga2024JCTC} second, a new potential energy surface was developed.\cite{Sunaga2025JCP_PES} Third, \emph{i.e.,} this work is about the development of property (dipole and electric polarizability) surfaces and their first application (and assessment) in vibrational transition computations in conjunction with a path-following Eckart frame.

\section{Theoretical and computational details\label{sec:theocomp}}
We first review the variational vibrational methodology (Sec.~\ref{sec:varvib}), and then, introduce the vibrational intensities of infrared and Raman spectra and corresponding molecular properties (Sec.~\ref{sec:IR_Raman}). Finally, we review the equivariant neural network method used in this work (Sec.~\ref{sec:NN_theory}).
%
%
\subsection{Variational vibrational computations \label{sec:varvib}}
\noindent 
We solve the vibrational Schrödinger equation 
\begin{align}
  \hat{H}_\text{vib} \Psi^\text{vib}_i = E_i \Psi^\text{vib}_i \;,
\end{align}
where the Hamiltonian, 
\begin{align}
  \hat{H}_\text{vib} 
  = 
  \hat{T}_\text{vib} + V \; ,
\end{align}
is the sum of the $\hat{T}_\text{vib}$ vibrational kinetic energy operator and the potential energy surface (PES).
For the potential energy part, we use the methanol PES of Ref.~\citenum{Sunaga2025JCP_PES} (PES2025), computed at the F12b-CCSD(T)/cc-pVTZ-F12 level and fitted with permutationally invariant polynomials.
As to the vibrational kinetic energy operator, it is first necessary to define physically-motivated internal (vibrational) coordinates, $\bos{\rho}=(\rho_1,\rho_2,\ldots,\rho_{D})$ with $D=3N-6$ for an $N$-atomic molecule. The coordinate transformation from the laboratory Cartesian coordinates to centre-of-mass translational, orientational, and vibrational coordinates is characterised by the (mass-weighted) metric tensor, $\bos{g}$. 

Following earlier work,\cite{Avila2019JCP_methodology,Avila2019JCP_CH4F-,Matyus2023CC} we write the vibrational kinetic energy operator in the following general form, 
\be
  \hat{T}_{\mathrm{vib}}
  =
  -\frac{\hbar^2}{2} 
  \sum_{k=1}^{D} \sum_{l=1}^{D} 
    G_{k l} \frac{\partial}{\partial \rho_k} \frac{\partial}{\partial \rho_l}
  -\frac{\hbar^2}{2} 
  \sum_{k=1}^{D} 
    B_k \frac{\partial}{\partial \rho_k}
  +U
\label{eq:keo}
\ee
with
\be
 B_k=\sum_{l=1}^{D} \frac{\partial}{\partial \rho_l} G_{l k}
 \label{eq:keo2}
\ee
and
\be
 U=\frac{\hbar^2}{32} \sum_{k=1}^{D}\sum_{ l=1}^{D}\left[\frac{G_{k l}}{\tilde{g}^2} \frac{\partial \tilde{g}}{\partial \rho_k} \frac{\partial \tilde{g}}{\partial \rho_l}+4 \frac{\partial}{\partial \rho_k}\left(\frac{G_{k l}}{\tilde{g}} \frac{\partial \tilde{g}}{\partial \rho_l}\right)\right] \; ,
 \label{eq:keo3}
\ee
and $D=12$ (for a six-atomic molecule), $\tilde{g}=\det\bos{g}$ and $G_{kl}=(\bos{g}^{-1})_{kl}$. 
We use the GENIUSH implementation of the numerical kinetic energy operator approach, and rely on the `$t$-vector formalism'~\cite{Matyus2009JCP,Matyus2023CC} to construct the kinetic energy coefficients (over the integration grid).

\subsubsection{Internal coordinates\label{sec:coord}}
\begin{figure}[htbp!]
  \centering
  \includegraphics[width=0.40\linewidth]{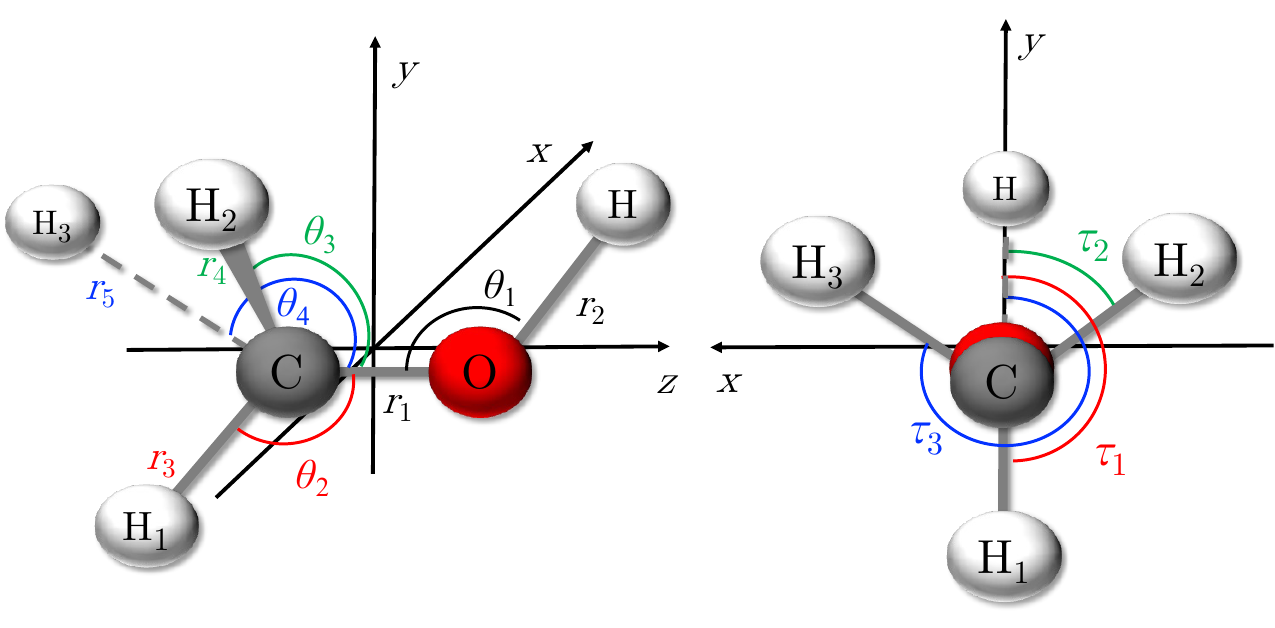}
  \caption{%
    Definition of the primitive internal coordinates and the primitive body-fixed (pBF) Cartesian frame of the methanol molecule used in this work. The pBF frame is shifted to the centre of mass.
  }
  \label{fig:methanol_cord}
\end{figure}

The vibrational coordinates of CH$_3$OH are defined as in previous work,\cite{Sunaga2024JCTC,Sunaga2025JCP_PES} following Refs.~\citenum{Meyer1969JCP_tau,Bell1994JMS_tau,Lauvergnat2014SA_methanol}. In short, we reiterate the coordinate definition for completeness.
First, `primitive' internal coordinates are defined similarly to the Z-matrix approach (Fig.~\ref{fig:methanol_cord}),
\bea\label{eq:cart_sym}
&\boldsymbol{r}_{\mathrm{C}}=  \mathbf{0}, \quad \boldsymbol{r}_{\mathrm{O}}=\left(\begin{array}{c}0 \\ 0 \\ r_1\end{array}\right), \\ \nonumber
&\boldsymbol{r}_{\mathrm{H}}=  \boldsymbol{r}_{\mathrm{O}}+\left(\begin{array}{c}0 \\ r_2 \sin \theta_1 \\ -r_2 \cos \theta_1 \end{array}\right), \\ \nonumber
&\boldsymbol{r}_{\mathrm{H}_1}=  \left(\begin{array}{c}-r_3 \sin \theta_2 \sin \tau_1 \\ r_3 \sin \theta_2 \cos \tau_1 \\ r_3 \cos \theta_2\end{array}\right), \\ \nonumber
&\boldsymbol{r}_{\mathrm{H}_2}=  \left(\begin{array}{c}-r_4 \sin \theta_3 \sin \tau_2 \\ r_4 \sin \theta_3 \cos \tau_2 \\ r_4 \cos \theta_3\end{array}\right), \\ \nonumber
&\boldsymbol{r}_{\mathrm{H}_3}=  \left(\begin{array}{c}-r_5 \sin \theta_4 \sin \tau_3 \\ r_5 \sin \theta_4 \cos \tau_3 \\ r_5 \cos \theta_4\end{array}\right) .
\eea
Then, these Cartesian coordinates are shifted so that the centre of mass is at the origin of the coordinate system, which completes the definition of the `primitive' body-fixed (pBF) frame used in this work. 

%
%

For describing the large-amplitude (torsional) motion, we use the linear combination of the $\tau_1,\tau_2,$ and $\tau_3$ dihedral angles that carry the permutational symmetry of the three methyl protons,\cite{Meyer1969JCP_tau,Bell1994JMS_tau,Lauvergnat2014SA_methanol}
\begin{align}
  \tau &=\frac{1}{3}\left(\tau_1+\tau_2+\tau_3\right) \; , \label{eq:tau} \\ 
  \varphi_1 &=\frac{1}{\sqrt{2}}\left(\tau_2-\tau_3\right) \; , \label{eq:vphi1} \\ 
  \varphi_2 &=\frac{1}{\sqrt{6}}\left(2 \tau_1-\tau_2-\tau_3\right) \; . \label{eq:vphi2}
\end{align}
The dynamically relevant interval for $\tau$ is its entire mathematical domain, $[0,2\pi)$. For $\varphi_1$ and $\varphi_2$, it is sufficient to use more restricted intervals,\cite{Sunaga2024JCTC,Sunaga2025JCP_PES} since they correspond to small-amplitude motions.

Then, for an efficient description of the small-amplitude vibrations, we define the $q_k, k=1,\ldots,D^\text{s}$ $(D^\text{s}=11)$ curvilinear normal coordinates,\cite{Avila2023PCCP_HCOOH,Daria2022JMS_HCOOH,Matyus2023CC,Sunaga2024JCTC,Sunaga2025JCP_PES} 
\be
  \Delta \xi_i=\sum_{k=1}^{D^\text{s}} \bar{L}_{i k} q_k 
  \quad\text{with}\quad 
  \Delta \xi_i=\xi_i- \xi_i^{\text {ref }}(\tau) \; ,
  \quad i=1,\ldots,D^\text{s}\; .
  \label{eq:defqk}
\ee 
In short,
$\bos{\xi}=(r_1,r_2,r_3,r_4,r_5,\theta_1,\theta_2,\theta_3,\theta_4,\varphi_1,\varphi_2)$ collects the primitive small-amplitude coordinates
and the $\bos{\xi}^{\text{ref}}(\tau)$ elements are the small-amplitude coordinates that provide the minimum energy (minimum-energy path, MEP) on the PES along the large-amplitude torsional motion, $\tau\in[0,2\pi)$. 
The $\bar{L}_{ik}$ linear combination coefficients are defined to minimise the coupling of the small-amplitude coordinates near the MEP, at the harmonic level, \emph{i.e.,}
$\bar{\bos{L}}$ diagonalizes the $\bar{\bos{G}}^\text{s} \bar{\bos{F}}^\text{s}$ matrix, where $\bar{\bos{G}}^\text{s}$ and $\bar{\bos{F}}^\text{s}$ are obtained by averaging the elements of the $D^\text{s}\times D^\text{s}$ $\bos{G}^\text{s}(\tau^{(i)})$ kinematic matrix and $\bos{F}^\text{s}(\tau^{(i)})$ force constant matrix over the three minimum structures, $\tau^{(i)} = \pi/3, \pi, 5\pi/3$. In this work, we use the MEP and the $\bar{L}_{ik}$ coefficients calculated in Ref.~\citenum{Sunaga2025JCP_PES}.
Then, the kinetic energy operator is constructed, Eqs.~\eqref{eq:keo}--\eqref{eq:keo3}, numerically over the integration grid, for the $q_k$ small-amplitude and the $\tau$ large-amplitude coordinates, collected in $\bos{\rho}=(q_1,\ldots,q_{D^\text{s}};\tau)$. 

The coordinates reviewed in this section allowed us to employ an efficient vibrational basis and integration-grid truncation techniques\cite{Avila2009JCP} for the small-amplitude motions, while retaining a non-truncated representation of the large-amplitude degree of freedom. \cite{Avila2019JCP_methodology,Avila2019JCP_CH4F-,Avila2023PCCP_HCOOH,Daria2022JMS_HCOOH}

\subsubsection{Vibrational Hamiltonian matrix using truncated basis and grid representations \label{sec:pruned}}
In this study, we employed a pruned product vibrational basis representation for the small-amplitude coordinates as in previous work. \cite{Avila2019JCP_methodology,Avila2019JCP_CH4F-,Avila2020PCCP,Papp2023MP,Daria2022JMS_HCOOH,Avila2023PCCP_HCOOH,Sunaga2024JCTC} 
The basis representation of the $i$th vibrational wave function, $\Psi^{\mr{vib}}_i$, is according to Eq.~41 of Ref.~\citenum{Sunaga2024JCTC},
\be\label{eq:non-DP_psi}
  \Psi^{\mr{vib}}_i\left(q_1, \ldots, q_{D^{\mathrm{s}}} ; \tau\right)  
  = 
  \sum_{f\left(n_1, \ldots, n_{D^\mathrm{s}}\right) \leq b} 
  \sum_{n_\tau=0}^{N_\tau} 
    C_{n_1, \ldots, n_{D^\mathrm{s}}, n_\tau}^i 
    \prod_{j=1}^{D^{\mathrm{s}}} \psi_{n_j}^{(q)}\left(q_j\right) \psi_{n_\tau}^{(\tau)}(\tau) \; ,
\ee
where the parameter $b$ will be used to refer to the basis set size and $f\left(n_1, \ldots, n_{D^\mathrm{s}}\right)$ is the simplest possible pruning function,
\be 
  f\left(n_1, \ldots, n_{D^\mathrm{s}}\right)
  =
  \sum_i^{D^\mathrm{s}} n_i \geq 0\; .
  \label{eq:bcond}
\ee
We use harmonic oscillator functions for $\psi_{n_j}^{(q)}\left(q_j\right)$ and Fourier functions for $\psi_{n_{\tau}}^{(\tau)}\left(\tau\right)$. 
The coordinate definition, Eq.~\eqref{eq:defqk}, which minimizes the coupling of the small-amplitude vibrations near the MEP, allows us to truncate the direct-product harmonic oscillator basis set used. This truncation is implemented by the `so-called' pruning function, Eq.~\eqref{eq:bcond}.

To construct the Hamiltonian matrix, potential energy and kinetic energy matrix elements must be computed. The action of the differential operators is calculated analytically, but integrating the coordinate-dependent functions (in the potential and kinetic energy) requires multidimensional numerical integration. In parallel with the basis truncation of the small-amplitude vibrations, it is possible to reduce the integration grid size (and avoid using rapidly growing direct-product grids) by the Smolyak quadrature scheme.\cite{Smolyak1963SSSR,Petras2003_Numer_Math,Avila2009JCP,Avila2011JCP,Avila2019JCP_methodology,Avila2019JCP_CH4F-,Avila2020PCCP}

In this work, the following basis and grid parameters are used in the GENIUSH-Smolyak computation
(detailed definition of the computational parameters can be found in Refs.~\citenum{Avila2019JCP_methodology,Daria2022JMS_HCOOH,Sunaga2024JCTC}): $n_\tau=32$ and $M_\tau=54$ for the basis and grid size of the large-amplitude motion; and $b=7$ and $H=21$ for the small-amplitude basis and grid truncation. In Ref.~\citenum{Sunaga2024JCTC}, it was numerically demonstrated that these parameters converge the vibrational energies within 0.5~$\mathrm{cm}^{-1}$\ up to 2000~$\mathrm{cm}^{-1}$\ from the zero-point vibrational energy (ZPVE) and within 1.5~$\mathrm{cm}^{-1}$\ up to 2200~$\mathrm{cm}^{-1}$. 
More sophisticated basis truncation conditions, beyond the simplest Eq.~\eqref{eq:bcond} relation, can be elaborated in the future for achieving better convergence of the higher-energy vibrations (up to the C-H, O-H stretching range), 
\emph{e.g.,} along the lines discussed in Refs.~\citenum{Avila2009JCP,Avila2011JCP,Daria2022JMS_HCOOH,Avila2023PCCP_HCOOH}.

\subsubsection{Transition dipole and polarizability integrals with the vibrational wave functions \label{sec:vibtrans}}
The variational vibrational wave functions are used to compute transition integrals with the dipole and polarizability components,
\be\label{eq:A_ij}
A_{ij}= \Braket{\Psi^{\mr{vib}}_i |\Vp | \Psi^{\mr{vib}}_j}.
\ee
In this work, $\Vp$ refers to the $\mu_a$ dipole moment component or the $\alpha_{ab}$ static polarizability element, where $a,b=x,y,z$ label the body-fixed frame (BF) axes. Technically, these integrals are calculated as potential-energy integrals. In practice, we used the generic matrix-vector multiplication procedure\cite{Avila2009JCP,Avila2011JCP, Avila2019JCP_methodology,Avila2019JCP_CH4F-} for transition property integration. So, we replaced the potential energy values with property values at the integration grid points.
In short, we write the property matrix element calculated by quadrature as,
\bea\label{eq:A_ij}
A_{ij}
&=& \sum_{n^{\prime}_{1} =0}^{n_{1} ^{\prime\max }} \sum_{k_1=1}^{k_1^{\max }} \psi_{n_1^{\prime}}^{(1)}\left(\rho_{1, k_1}\right) \\ \nonumber
&\ldots& \sum_{n^{\prime}_{D} =0}^{n_D ^{\prime\max }} \sum_{k_D=1}^{k_D^{\max }} \psi_{n_D^{\prime}}^{(D)}\left(\rho_{D, k_D}\right) 
C^i_{n^{\prime}_1, \ldots, n^{\prime}_{D}} W_{k_1, \ldots, k_D}^{\mathrm{S}} \Vp \left(\bos{\rho}_{k_1, \ldots, k_D}\right) \\ \nonumber
&\times& 
\sum_{n_D=0}^{n_D^{\max }} \psi_{n_D}^{(D)}\left(\rho_{D, k_D}\right) \ldots \sum_{n_1=0}^{n_1^{\max }} \psi_{n_1}^{(1)}\left(\rho_{1, k_1}\right) C^j_{n_1, \ldots, n_{D}} \\ \nonumber
&=&
\sum_{n^{\prime}_{1} =0}^{n_{1} ^{\prime\max }} \ldots \sum_{n^{\prime}_{D} =0}^{n_{D} ^{\prime\max }} C^i_{n^{\prime}_1, \ldots, n^{\prime}_{D}} f_{n^{\prime}_1, \ldots, n^{\prime}_{D}}^{\mr{MVP}}(\Vp,C^j_{n_1, \ldots, n_{D}}),
\eea
where $W^{\mr{S}}$ is the weight for the numerical integration. 
In the last step, it is indicated that a matrix-vector multiplication is calculated first with
\bea
&&f_{n^{\prime}_1, \ldots, n^{\prime}_{D}}^{\mr{MVP}}(\Vp,C^j_{n_1, \ldots, n_{D}}) \\ \nonumber
&=&
\sum_{k_1=1}^{k_1^{\max }} \psi_{n_1^{\prime}}^{(1)}\left(\rho_{1, k_1}\right) \ldots \sum_{k_D=1}^{k_D^{\max }} \psi_{n_D^{\prime}}^{(D)}\left(\rho_{D, k_D}\right) 
W_{k_1, \ldots, k_D}^{\mathrm{S}} \Vp \left(\bos{\rho}_{k_1, \ldots, k_D}\right) \\ \nonumber
&\times&
\sum_{n_D=0}^{n_D^{\max }} \psi_{n_D}^{(D)}\left(\rho_{D, k_D}\right) \ldots \sum_{n_1=0}^{n_1^{\max }} \psi_{n_1}^{(1)}\left(\rho_{1, k_1}\right) C^j_{n_1, \ldots, n_{D}} \; . 
\eea
In the above expressions, we simplify notation by not explicitly distinguishing indices for the small-amplitude and the large-amplitude vibrations: $\{n_1,n_2,\ldots,n_D\}\equiv\{n_1,n_2,\ldots,n_{D_\mr{s}},n_{\tau}\}$ (for basis functions) and $\{k_1,k_2,\ldots,k_D\}\equiv\{k_1,k_2,\ldots,k_{D_\mr{s}},k_{\tau}\}$ (for quadrature grid points).
$\bos{\rho}_{k_1, \ldots, k_D}$ is the compact notation of $\left\{\rho_{k_1},\rho_{k_2},\ldots, \rho_{k_D}\right\}$.

\subsubsection{Path-following Eckart frame for the property tensor computations \label{sec:pathEckart}}
The primary aim of this work is the development of the electric dipole and polarisability surfaces for the methanol molecule, followed by their first application and assessment. Hence, we restrict our computations to vibrational transition integrals (and simulated spectra) which depend on the actual body-fixed (BF) frame definition (in which the property vector and tensor elements are computed).

The vibrational band profile can often be well approximated by using the (frame-dependent) vibrational transition integrals, instead of the rigorous, frame-independent rovibrational transition integrals. 
The vibrational approximation uses the Eckart frame for semi-rigid molecules~\cite{LeSueur1992MP_Eckart_transition} (or molecules with structural isomers, \emph{e.g.}, Ref.~\citenum{Avila2023PCCP_HCOOH}), which minimises rovibrational coupling and allows approximating the rovibrational wave function by the product of vibrational and rigid rotor wave functions. Since the methanol molecule has one large-amplitude motion (LAM), we use a LAM-following frame definition following Ref.~\citenum{Lauvergnat2016JCP}.

The orientation of the Eckart frame (for a single reference structure) is defined by the rotational Eckart condition,\cite{Eckart1935PR} \emph{i.e.,} the Eckart-frame $\boldsymbol{r}_i$ Cartesian coordinates must fulfil,
\be\label{eq:Eckart}
  \sum^N_{i=1} m_i \boldsymbol{a}_i \times \boldsymbol{r}_i=0 \; ,
\ee
with $m_i$ masses associated with the nuclei and the $\boldsymbol{a}_i$ reference (most often, the equilibrium) structure of the molecule. 

Since in methanol, there is a single large-amplitude ($\tau$) degree of freedom, it is straightforward to define a path-following Eckart frame,
\be\label{eq:path-following_Eckart}
  \sum^N_{i=1} m_i \boldsymbol{a}_i(\tau) \times \boldsymbol{r}_i=0 \; ,
\ee
where we choose the minimum energy path (Sec.~\ref{sec:coord}) to define the $\boldsymbol{a}_i(\tau)$ reference structure for every $\tau$ value.
Alternatively, a more sophisticated frame definition for molecules with one LAM is provided by the Eckart-Sayvetz condition,\cite{Sayvetz1939JCP,Szalay2014JCP_Sayvetz} which has been applied in rovibrational computations of the ammonia molecule.\cite{Yurchenko2005MP_Sayvetz}

At this point, we reiterate technical details of constructing the $\boldsymbol{a}_i(\tau)$ Cartesian structures along the MEP.\cite{Sunaga2024JCTC,Sunaga2025JCP_PES} 
First, the $\xi_k^\mr{ref}(\tau_n)$ $(k=1,\ldots,11)$ `MEP' values of the internal coordinates were obtained by minimization of the PES at a series of $\tau_n=(n-1) 6^{\circ} \;(n=1, \ldots, 61)$ values of the torsional coordinate; next, the computed values of $\xi_k^\mr{ref}(\tau_n)$ $(k=1,\ldots,11)$ were interpolated by cubic splines. The Cartesian coordinates (for the path-following reference structure) are calculated according to Eqs.~\eqref{eq:cart_sym}--\eqref{eq:vphi2}. Then, for an arbitrary value of the internal coordinates $(\bos{\rho}=[\bos{\xi};\tau]$ (a quadrature point used for the multi-dimensional integration, Sec.~\ref{sec:pruned}), first, the primitive Cartesian coordinates are calculated according to Eqs.~\eqref{eq:cart_sym}--\eqref{eq:vphi2}, and then, the orientational Eckart condition is solved using 
the approach based on quaternion algebra~\cite{Krasnoshchekov2014JCP_Eckart} as implemented in Ref.~\citenum{FaMaCs14}.

Instead of interpolation, one could use a simple switching function that smoothly connects the reference geometries along the torsional coordinate~\cite{Lauvergnat2016JCP} or define a (physically-motivated) analytic expression for $\boldsymbol{a}_i(\tau)$ for the fixed SAM coordinates along the LAM.\cite{Yurchenko2005MP_Sayvetz}

%
%
\subsection{Infrared and Raman intensity calculations}\label{sec:IR_Raman}
\subsubsection{Integrated absorption coefficient for infrared transitions}
In this section, we briefly reiterate details for the integrated absorption coefficient,\cite{{Wilson1955book,Craig1998_Molecular_quantum_electrodynamics,Brown1998_mol_spec,Neugebauer2002JCC,Bunker2006_molsym}} which is computed in this work for the methanol molecule. 
A molecular-specific transition property, the $A^{\mr{IR}}$ integrated absorption coefficient, can be expressed  by
\be\label{eq:A_IR_T_depend}
A^{\mr{IR}}_{\text{fi}} = 
\frac{8 \pi^3 N_A \tilde{\nu}_{\mr{fi}} \exp \left(-hc\tilde{\nu}_{\mr{i}} / k T\right)\left[1-\exp \left(-h c \tilde{\nu}_{\mr{fi}} / k T\right)\right]}{\left(4 \pi \epsilon_0\right) 3 h c Q} S(\mr{f} \leftarrow \mr{i}) \; ,
\ee
where $N_A$ is Avogadro constant, $\tilde{\nu}_{\mr{fi}} = \tilde{\nu}_{\mr{f}} - \tilde{\nu}_{\mr{i}}$ is the energy difference between the final and initial state, $k$ is the Boltzmann constant, $T$ is the absolute temperature (assuming molecules are in thermal equilibrium), $h$ is the Plank constant, $c$ is the speed of light in vacuum, and $\varepsilon_0$ is the permittivity of vacuum. $Q$ is the partition function,
$Q=\sum_w g_w \exp \left(-\tilde{\nu}_{w} / k T\right)$ with
the $g_w$ degeneracy factor and the $\tilde{\nu}_{w}$ energy of the $w$ state.
In the electric dipole approximation, the $S(\mr{f} \leftarrow \mr{i})$ line intensity can be calculated from the electric dipole transition integral  
\be
  S(\mr{f} \leftarrow \mr{i})
  =
  \left\langle\mu_\mr{\mr{f}\mr{i}}\right\rangle^2 \; .
\ee

In the low-temperature limit of Eq.~\eqref{eq:A_IR_T_depend}, which will be used in this work and \emph{e.g.,} relevant to supersonic jet IR experiments, $A^\mr{IR}$ simplifies to
\be
A^{\mr{IR}}  =  \frac{N_A   \pi}{3 \varepsilon_0 \hbar c} \tilde{\nu}_\mr{f i}S(\mr{f} \leftarrow \mr{i}) \; . 
\ee
Using the CODATA2022 conversion factors, the final working expression is
\be
A^{\mr{IR}}/(\mr{km} \; \mr{mol}^{-1}) = 2.50664320 \  \nu_\mr{f i}/\mr{cm}^{-1} \left\langle\mu_\mr{f i}\right\rangle^2/\mr{Debye}^2 \; ,
\label{eq:A_IR}
\ee
and we use the approximation
\be\label{eq:mu^2}
\left\langle \mu_{{\mr{f}}{\mr{i}}} \right\rangle^2 
\approx 
  \sum_{a=x,y,z} 
  |\braket{\Psi^{\mr{vib}}_{\mr{f}}|\mu^\text{BF}_a|\Psi^{\mr{vib}}_{\mr{i}}}|^2 \; ; 
\ee
where $\mu^\text{BF}_a$ is the electric dipole moment (from electronic structure theory and fitting), and the approximation refers to the vibrational approximation to the transition dipole moment, by exploiting the approximate factorisation of the rovibrational wave function.

%
%
\subsubsection{Raman intensity} 
The Raman intensity depends on the relative orientation of the observation axis and the polarisation of the incident light's electric vector. In what follows, a short collection of working formulae is provided. The low-temperature limit is assumed and any higher-order, \emph{e.g.,} hyperpolarizability contributions are neglected.
The parallel and perpendicular orientations of the scattered intensities (Sec.~3.6 of Ref.~\citenum{Wilson1955book}), being $X$ the observational direction, are
\begin{align}
\label{eq:I_Raman_average}
(\mr{A})\;\; I^{\mr{R}}(\mr{obs.} \parallel)&=
\frac{\pi^2 c \tilde{v}^4}{2 \varepsilon_0}
\left(\braket{\alpha^2_{Y X}}_{\mr{iso}} + \braket{\alpha^2_{Z X}}_{\mr{iso}}\right)
\mathcal{E}_0^2\; ; 
\quad \mathcal{E}_X = \mathcal{E}_0 
\nonumber
\\ 
(\mr{B})\;\; I^{\mr{R}}(\mr{obs.} \perp) &=
\frac{\pi^2 c \tilde{v}^4}{2 \varepsilon_0}
\left(\braket{\alpha^2_{Y Z}}_{\mr{iso}} + \braket{\alpha^2_{Z Z}}_{\mr{iso}}\right)
\mathcal{E}_0^2\; ; 
\quad \mathcal{E}_Z = \mathcal{E}_0 \; .
\end{align}
$\alpha_{AB}$ is the transition polarizability, where $A,B=X,Y,Z$ are the laboratory-frame coordinates and 
$\mathcal{E}_0$ is the amplitude of the electric field vector, $\vec{\mathcal{E}}=(\mathcal{E}_X,\mathcal{E}_Y,\mathcal{E}_Z)$. $\braket{\ldots}_{\mr{iso}}$ is the isotropic average, taking the free rotation of the molecule into account. 
The sum of the polarizability tensor components can be expressed in terms of the rotational-invariant Placzek invariants, $\mathscr{G}^{(i)}$,\cite{Placzek1931EurPhysJA,Long2002_Raman}
\be
\mathscr{G} = \sum_{ab}|\alpha_{ab}|^2 = \mathscr{G}^{(0)} + \mathscr{G}^{(1)} + \mathscr{G}^{(2)}, 
\ee
where
\begin{align}
\label{eq:a_gamma} 
\mathscr{G}^{(0)} = 3a^2 &= \frac{1}{3}\left\{\left|\alpha_{x x}+\alpha_{y y}+\alpha_{z z}\right|^2\right\} \\ \nonumber
\mathscr{G}^{(1)}= \frac{2}{3}\delta^2 &= \frac{1}{2}\left\{\left|\alpha_{x y}-\alpha_{y x}\right|^2+\left|\alpha_{x z}-\alpha_{z x}\right|^2+\left|\alpha_{y z}-\alpha_{z y}\right|^2\right\} \\ \nonumber
\mathscr{G}^{(2)}= \frac{2}{3}\gamma^2 &= \frac{1}{2}\left\{\left|\alpha_{x y}+\alpha_{y x}\right|^2+\left|\alpha_{z x}+\alpha_{x z}\right|^2+\left|\alpha_{y z}+\alpha_{z y}\right|^2\right\} \\ \nonumber
&+ \frac{1}{3}\left\{\left|\alpha_{x x}-\alpha_{y y}\right|^2+\left|\alpha_{x x}-\alpha_{z z}\right|^2+\left|\alpha_{y y}-\alpha_{z z}\right|^2\right\} \; .
\end{align}
The quantities $a$, $\delta$, and $\gamma$ are called the mean polarizability, antisymmetric anisotropy, and the anisotropy of the polarizability, respectively. 

Similarly to the dipole transition integrals, we use the vibrational approximation (approximate factorization of the rovibrational wave function), and in what follows, we compute vibration-only quantities.
First, we evaluate the $\alpha_{ab}$ polarizability transition integral (with the vibration-only wave functions and the polarizability in a body-fixed frame)
\be\label{eq:alpha}
\alpha_{ab}
=
\braket{\Psi^{\mr{vib}}_\mr{f}|\alpha_{ab}^\text{BF}|\Psi^{\mr{vib}}_\mr{i}} \; .
\ee
Then, similar to earlier work,\cite{Avila2023PCCP_HCOOH} perpendicular, parallel, and total vibrational Raman activities are calculated according to
\begin{align}\label{eq:A_Raman}
  A^\mr{R}_\perp
  &=
  45\left\langle\alpha_{YX}^2\right\rangle_{\mr{iso}} = 45\left\langle\alpha_{ZX}  ^2\right\rangle_{\mr{iso}} =
  3{\gamma^2}, \\ \nonumber
  A^\mr{R}_\parallel
  &= 
  45\left\langle\alpha_{ZZ}^2\right\rangle_{\mr{iso}} =
  45 a^2+4 \gamma^2. \\ \nonumber
  A^\mr{R}
  &= 
  45\left(\left\langle\alpha_{YZ}^2\right\rangle_{\mr{iso}}+\left\langle\alpha_{ZZ}^2\right\rangle_{\mr{iso}} \right)  = 
  45 a^2+7 \gamma^2 \;.   
\end{align}
The depolarization ratio, which is the ratio between the perpendicular and parallel components, is defined by
\be
\rho = \frac{A^\mr{R}_\perp}{A^\mr{R}_\parallel} = \frac{3\gamma^2}{45a^2+4\gamma^2} \; .
\ee
This quantity ($0< \rho\leq 0.75$) is useful to know the isotropic and anisotropic contributions in the Raman transition.

\subsection{Representation of first- and second-rank tensorial functions of the nuclear coordinates with equivariant neural networks}\label{sec:NN_theory}
\noindent %
To compute the vibrational transition integrals for the electric dipole and polarizability, we must evaluate these quantities over the entire integration grid (Sec.~\ref{sec:varvib}). Instead of performing direct \emph{ab initio} computations at all grid points, we construct property surfaces fitted to a smaller number of $\emph{ab initio}$ points. \\

\subsubsection{Property representations}
Representations of atomic environments allow encoding geometric information in a manner that obeys desired physical symmetries, enabling efficient fitting of properties that are functions of atomic coordinates.
It is typically expected that properties remain invariant under permutations of equivalent atomic nuclei and rigid translations, and transform equivariantly under rigid rotations.
For example, scalar functions of the nuclear coordinates, such as the potential energy or the isotropic component of the polarisability, are invariant to rotations.
Vector quantities, such as the dipole moment $\boldsymbol{\mu}$, are expected to transform as
\[
\boldsymbol{\mu} \to \mathbf{R} \boldsymbol{\mu}
\]
if the nuclear coordinates $\{\mathbf{r}_i\}$ are rigidly rotated as $\{\mathbf{R}\mathbf{r}_i\}$ where $\mathbf{R}$ is a rotation matrix.
Higher rank tensor quantities follow suit, such as the polarisability tensor $\boldsymbol{\alpha}$ which transforms according to $\mathbf{R}\boldsymbol{\alpha}\mathbf{R}^T$.
In order to fit such quantities efficiently and to ensure exact equivariance of the surrogate model predictions, it is desirable to employ an Ansatz where the outputs obey the required symmetry behavior by construction.

Generally, in neural networks input features undergo successive non-linear transformations, losing symmetry properties in the process.
In contrast, equivariant neural networks\cite{geiger2022e3nneuclideanneuralnetworks} combine input features with each other and with adjustable weights to generate new features strictly obeying rotational equivariance of arbitrary rank tensor quantities.
It is therefore possible to extract final predictions which obey exact equivariance with respect to rotations, and by adjusting the free weights using backpropagation the model can be fitted to reproduce the training data.

We modelled the total dipole moment $\boldsymbol{\mu}$ and polarisability $\boldsymbol{\alpha}$ of molecule $M$ as sums of atomic contributions
\bea
\boldsymbol{\mu}_M = \sum_{i\in M} \boldsymbol{\mu}(\mathcal{X}_i)
\\ \nonumber
\boldsymbol{\alpha}_M = \sum_{i\in M} \boldsymbol{\alpha}(\mathcal{X}_i)
\eea
where $\mathcal{X}_i$ represents the atomic environment of atom $i$.
Such decomposition of molecular properties into sums of local contributions has been a very common choice since the dawn of interatomic potentials\cite{Lennard-Jones_1931}, which sought to find the total energy as a sum of atomic or bond energies.
While not quantum mechanically observable quantities, local contributions are justified by the concept of nearsightedness of electronic matter\cite{Prodan.2005}, and form a convenient framework to model extendable systems.
In our case, this choice ensures that in (the hypothetical) case when molecule $M$ is dissociated into two non-interacting fragments $M_1$ and $M_2$, the individual dipoles and polarisabilities of $M_1$ an d $M_2$ can be recovered independently.
The functions $\boldsymbol{\mu}(\mathcal{X})$ and $\boldsymbol{\alpha}(\mathcal{X})$ are constructed such that they transform as first and second-rank tensors, respectively, as the environment $\mathcal{X}$ is rotated.
It follows that the total dipole and polarisability tensors transform accordingly.

It is useful to reiterate that Cartesian tensors can be decomposed into spherical tensors.
The dipole moment can be represented in the $L=1$ spherical basis, while the polarisability is a direct sum of $L=0$ (rotationally invariant) and $L=2$ representations.
Using equivariant descriptors, including adjustable parameters, of atomic environments that also transform according to $L=0,1,2$ allows us to fit dipole moments and polarisabilities.

In this work, we employed the MACE\cite{Batatia2022_proc_MACE} framework, which was used before to model the dipole and polarisability of water.\cite{10.1039/d3fd00113j}
As our focus remains the isolated methanol molecule, we found it sufficient to only include geometric information in the local atomic representation $\mathcal{X}_i$ and it remains practical to include all atoms as neighbors.
In MACE, the environment of atom $i$ is first defined by the tuples of atomic neighbor vectors and species $\{(\mathbf{r}_{ij}, Z_j)\}_{j\in \mathcal{X}_i}$, from which the features $\mathbf{A}_i$ are obtained by
\be
A_{i,klm}=\sum_{j \in \mathcal{X}_i} R_{kl}(r_{ij}) Y_{lm} (\hat{\mathbf{r}}_{ij}) W_{kZ_j}
\textrm{,}
\ee
where $R_{kl}$ are radial basis function, $Y_{lm}$ are the spherical harmonics functions and $\mathbf{W}$ are the species adjustable embedding weights\cite{Drautz2019PRB_ACE}.
The MACE architecture then proceeds to repeatedly (i) compose direct products of the features, including adjustable weights to achieve a desired body order representation $\nu$; 
(ii) aggregate features of neighboring atoms in a message passing operation, supplying a rich set of expressive, equivariant features.
In the last step, non-linear operations, in the form of gates, are applied to invariant features ($L=0$), and the direct product of these are formed with $L=1$ and $L=2$ equivariant features, preserving the equivariance.
We note that while message passing (referred as number of interactions in MACE) increases the receptive field of the model by transferring information of neighbors of neighbors successively, it is not strictly necessary here, as a moderate cutoff radius already encompasses all atoms in the molecule in case of most data points.
Instead, we use message passing to increase the expressivity of our representation, noting that body order and message passing  are interlinked\cite{Nigam.2022}.
The fitting process minimises the loss function obtained from the training labels -- dipole vector and polarisability matrix elements -- and the corresponding predicted properties by adjusting the weights in the MACE network.

\begin{table}[htbp!]
\caption{
Key input parameters of MACE employed in the property surface fitting.
}\label{tbl:MACE_param}
\centering
\begin{tabular}{@{} llc @{}}
\hline\hline
Parameter & \multicolumn{1}{c}{Description} & Value\tabularnewline
\hline
$r_\mr{cut}/$\AA & cutoff radius & 5\tabularnewline
$T$ & \# of interactions & 2\tabularnewline
$\nu$ & correlation order & 3\tabularnewline
$L$ & Max L & 2\tabularnewline
$k$ & channel & 32\tabularnewline
$B$ & batch size & 32\tabularnewline
$N_\mr{epoch}$ & \# of epoch & 20\tabularnewline
\hline\hline
\end{tabular}
\end{table}

\subsubsection{Electric dipole and polarizability ab initio computations \& fitting }\label{sec:surface}
\paragraph{Ab initio computations}
The DALTON program package~\cite{daltonpaper,Dalton2024} was used for all electronic property computations. For the electric dipole moment, we employed the coupled-cluster singles and doubles (CCSD) method~\cite{Halkier1997JCP_CCSD,Hald2003JCP_CCSD} including the orbital relaxation effects. For the static electric polarizability, we employed the second-order polarisation propagator approximation (SOPPA) with the CCSD method.\cite{Sauer1997JPB_SOPPA} The aug-cc-pVTZ basis sets~\cite{Dunning1989JCP_ccpv,Kendall1992JCP_aug_ccpv} were used in all property computations. The differences (for the equilibrium geometry) at the aug-cc-pVTZ and aug-cc-pVDZ are less than 1.5\% and 0.5\% for dipole moment and polarizability, respectively.

%
%
\begin{table*}[htbp!]
\caption{%
  Assessment of the quality of the dipole moment and polarizability surfaces of methanol developed in this work. The pBF frame (Sec.~\ref{sec:coord}) was used.
  RMSE: Root-Mean-Squared-Error,  MAPE: Mean-Absolute-Percentage Error (in \%). The errors for the training and test sets are listed.
  The RMSE and the equilibrium values (Eq.) are in $e^2a^2_0E^{-1}_\mr{h}$ for the polarizability and $ea_0$ for the dipole moment. 
}\label{tbl:error_property_PES}
\centering
\begin{tabular}{@{} ll rrr rrr rrr @{}}
\hline\hline\\[-0.255cm]
 &  & \multicolumn{6}{c}{polarizability ($\alpha$)} & \multicolumn{3}{c}{dipole moment ($\mu$)}\tabularnewline
 &  & $xx$ & $yy$ & $zz$ & $xy$ & $yz$ & $xz$ & $x$ & $y$ & $z$\tabularnewline
\hline\\[-0.255cm]
Eq. & \textit{ab} \textit{initio} 
                    & 19.649 & 20.261 & 23.130 &  0.000  & 0.577 &  0.000 &  0.0000 &  0.5417 &  --0.3930 \tabularnewline 
Eq. & prop. surface & 19.650 & 20.253 & 23.106 &  0.000  & 0.554 &  0.000 &  0.0000 &  0.5414 & --0.3929 \tabularnewline
\hline\\[-0.25cm]
RMSE      & training & 0.048 & 0.052  & 0.079  & 0.029   & 0.046 & 0.032  &  0.0018 & 0.0028 & 0.0028\tabularnewline
          & test     & 0.049 & 0.057  & 0.080  & 0.028   & 0.046 & 0.030  &  0.0018 & 0.0026 & 0.0031 \\[1mm]
MAPE      & training & 0.15  & 0.16   & 0.16   & $^\ast$27.11 & 19.06 & $^\ast$18.68 & $^\ast$42.95 & 0.41 & 1.34\tabularnewline
          & test     & 0.16  & 0.16   & 0.16   & $^\ast$19.79 & 29.20 & $^\ast$16.00 & $^\ast$20.86 & 0.40 & 1.05 \\[1mm]
\hline\hline
\end{tabular}
~\\[2mm]
$^\ast$~Regarding the large mean-absolute-percentage errors (MAPE), we note the smallness of the absolute value of the component, \emph{e.g.,} at the configurations close to the equilibrium structure.
\end{table*}

\paragraph{Fitting details}
We develop the dipole moment (vector) and polarizability (second-rank tensor) surfaces using a neural-network-based fitting program, called MACE.\cite{Batatia2022_proc_MACE}
In MACE, equivariant descriptors are combined to result in a global feature transforming as $L=1$, corresponding to the dipole vector target.
The direct sum of features transforming as $L=0$ and $L=2$ were used to fit the polarizability tensor, with $L=0$ features corresponding to the isotropic component and $L=2$ corresponding to the traceless and symmetric part.
The dipole moment and polarizability surfaces were trained independently resulting in separate models without weight sharing; that is, polarizabilities did not contribute to the loss when the dipole moment surface was fitted (and \emph{vice versa}). 
The training was performed using the 35~000 geometries, which were randomly selected from the 39~401 geometry points used for fitting PES2025~\cite{Sunaga2025JCP_PES} after removing nine distorted geometry points. The associated test set is what remains after removing the 35~000 geometry points and distorted geometry points from the 39~401 geometry points (4392 geometry points). Although more geometry points were generated during the PES development stage (`spares set', see Ref.~\citenum{Sunaga2025JCP_PES}), the 35~000 geometry points are sufficient according to the convergence test performed in the \SM.
Further details are provided in Sec S1 of \SM.
The MACE parameters employed for the fitting are summarized in Table~\ref{tbl:MACE_param}. Except for the number of batches and epochs, we employed the default parameters of MACE and similar to those used by others\cite{10.1039/d3fd00113j}.
The validation loss was converged well within the epochs. 
The parameter dependence on the number of baches and correlation order is small (Tables~S3 and S4 of \SM).
The advantages of MACE compared with the traditional polynomial expansion are as follows: 
a) the fitting functions are functions of the Cartesian coordinates of the nuclei, and by construction, they have the exact spatial rotational properties \cite{Bartok2010arXiv_rot_invariant,Batatia2022_proc_MACE} of a 1st- and  2nd-rank tensor (here, used for the dipole vector and the polarizability tensor, respectively); 
b) the appearance of holes and the possibility of overfitting is attenuated by regularization techniques commonly used when fitting neural networks, \emph{e.g.,} the use of stochastic gradient descent, weight decay~\cite{Guodong2018arXiv_weight_decay} and exponential moving average (EMA)~\cite{Morales-Brotons2024arXiv_EMA}; and 
c) the flexibility of the MACE model allows excellent representation of the target properties throughout the training set which includes configurations far from the equilibrium, while only requiring a moderate number of training geometries, as shown in Figs.~S1--S2 of the \SM.

Table~\ref{tbl:error_property_PES} provides details regarding the quality of the fitted property surfaces. The data used for this table was calculated in the primitive body-fixed frame (Sec.~\ref{sec:coord}).
Some components ($\mu_{x}$ and $\alpha_{ij}\;(i\neq j)$) show rather large mean-absolute-percentage errors (MAPEs). However, the root-mean-square errors (RMSEs) of all components are of a similar order of magnitude and are much smaller than those of the dominant component values at the equilibrium structure for each property ($\mu_{y}$, $\mu_{z}$, and $\alpha_{ii}$). 
The \emph{ab initio} values at the equilibrium structure (the three equilibrium structures are permutationally equivalent) are well reproduced for both property surfaces.

\section{Numerical results}\label{sec:result}

\subsection{Electric dipole and polarizability surfaces of the methanol molecule}
\begin{figure*}[htbp!]
    \centering
    \includegraphics[width=0.9\linewidth]{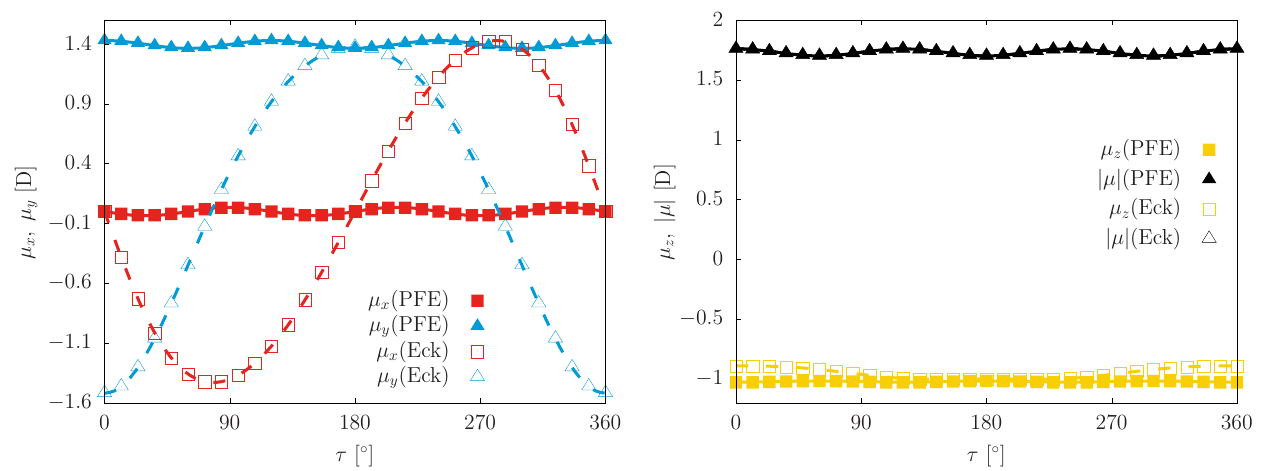}
    \caption{%
      Electric dipole moment of CH$_3$OH vs. $\tau$: components and $|\mu|$ length, as a function of the torsional angle; the other 11 internal coordinates were relaxed (MEP) on PES2025.\cite{Sunaga2025JCP_PES} 
      PFE: path-following Eckart frame, Eck: (single-reference) Eckart frame with the reference structure orientation according to pBF (Sec.~\ref{sec:coord}).
      The curves show the property surface components (and length), the points are the \emph{ab initio} values (at the same CCSD/aug-cc-pVTZ level of theory as used for the training dataset). 
    }
    \label{fig:1D_tor_mu}
\end{figure*}

\begin{figure*}[htbp!]
    \centering
    \includegraphics[width=1.0\linewidth]{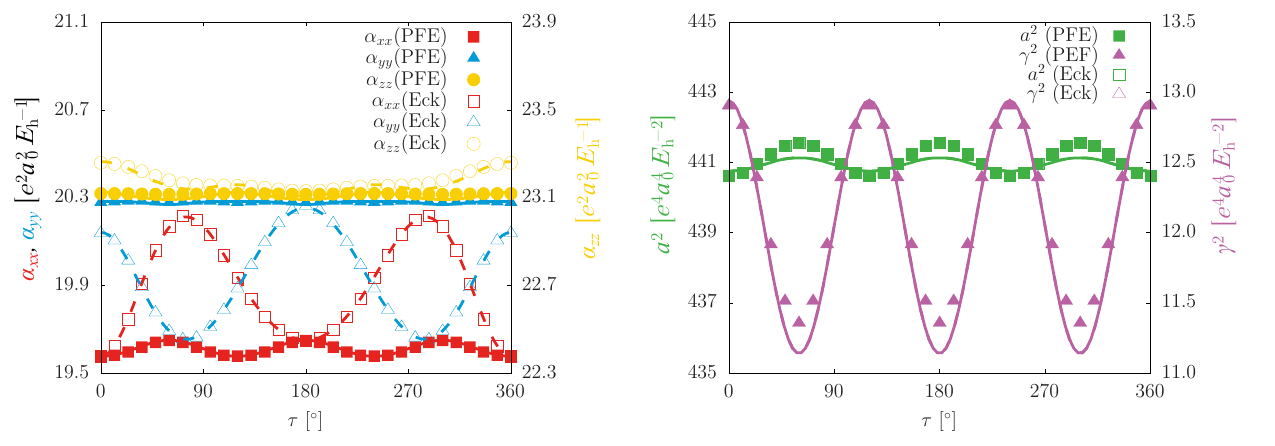}
    \caption{
      Polarizability of CH$_3$OH vs. $\tau$: components and squared mean (isotropic) and anisotropic polarizabilities, as a function of the torsional angle with the other 11 internal coordinates relaxed (MEP) on PES2025.\cite{Sunaga2025JCP_PES}
      PFE: path-following Eckart frame, Eck: (single-reference) Eckart frame with the reference structure orientation according to pBF (Sec.~\ref{sec:coord}).
      The curves show the property surface, the points are the \emph{ab initio} values (at the same CCSD/aug-cc-pVTZ level of theory as the training dataset). 
    }
    \label{fig:1D_tor_pol}
\end{figure*}

Figures~\ref{fig:1D_tor_mu} and~\ref{fig:1D_tor_pol} show the electric dipole moment and the polarizability, respectively, as a function of the torsion angle.
In the figures, the property components are calculated in the path-following Eckart frame, Eq.~\eqref{eq:path-following_Eckart}, as well as in the single-reference Eckart frame, Eq.~\eqref{eq:Eckart} (corresponding to the $\tau=180^\circ$ equilibrium structure). The reference structure is oriented according to Fig.~\ref{fig:methanol_cord} (Sec.~\ref{sec:coord}).
The figures show that the property surfaces correctly reproduce the periodicity of the \textit{ab} \textit{initio} values (also shown in the figures). 

Regarding the dipole moment (Fig.~\ref{fig:1D_tor_mu}), we observe that $\mu_x$ is antisymmetric with respect to the $\tau$ torsion, while $\mu_y$ and $\mu_z$ are symmetric to it. This behaviour is due to the choice of the reference structure, in which the C-O-H unit is in the $yz$ plane (Fig.~\ref{fig:methanol_cord}).

Regarding the polarizability surface (Fig.~\ref{fig:1D_tor_pol}), we see an overall good agreement for the surface and the \emph{ab initio} points. 
A comment regarding the square of the mean polarizability, $a^2$, is necessary. The deviation of the fitted surface and the \emph{ab initio} points may seem large at first glance, but in fact the relative deviation is only 0.1~\%, and $a^2$ is almost constant with respect to $\tau$. 

A somewhat larger deviation, but still only 2~\% on the relative scale, is seen for the squared anisotropic polarisability, $\gamma^2$, at the equilibrium structures ($\tau=60,180,300^\circ$). This can be attributed to (a small) underestimation of $\alpha_{yz}$ and $\alpha_{zz}$  by the fitted surface. 
%
All in all, the largest relative error of $\gamma^2$ (2~\%) is still comparable to the estimated error of the electronic structure method (CCSD/aug-cc-pVTZ), which is at least several percent.

%
%

\begin{figure*}[htbp!]
  \centering
  \includegraphics[width=1.0\linewidth]{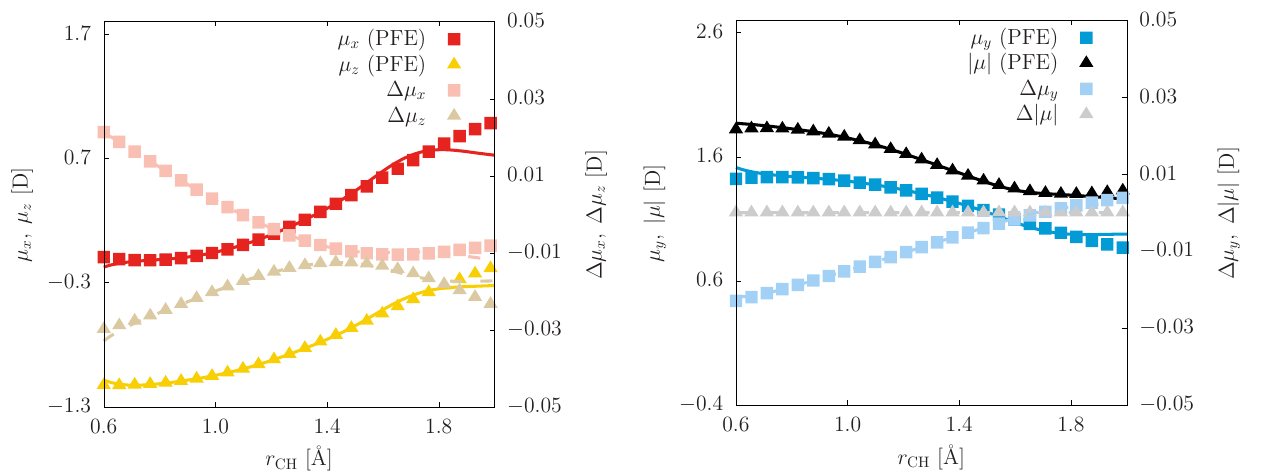}
  \caption{%
      Electric dipole moment of CH$_3$OH vs. $r_4$: components and $|\mu|$ length, as a 
      function of the $r_4$ \ce{CH} bond distance with all other coordinates fixed at one of the equilibrium structures of PES2025.\cite{Sunaga2025JCP_PES}
      PFE: path-following Eckart frame. 
      $\Delta \mu = \mu(\mr{pBF}) - \mu(\mr{PFE})$, where pBF refers to the primitive body-fixed frame (Sec.~\ref{sec:coord}).
      The curves show the property surface, the points are the \emph{ab initio} values (at the same CCSD/aug-cc-pVTZ level of theory as the training dataset).   
  }
    \label{fig:r4_CH_mu}
\end{figure*}

\begin{figure*}[htbp!]
  \centering
  \includegraphics[width=1.0\linewidth]{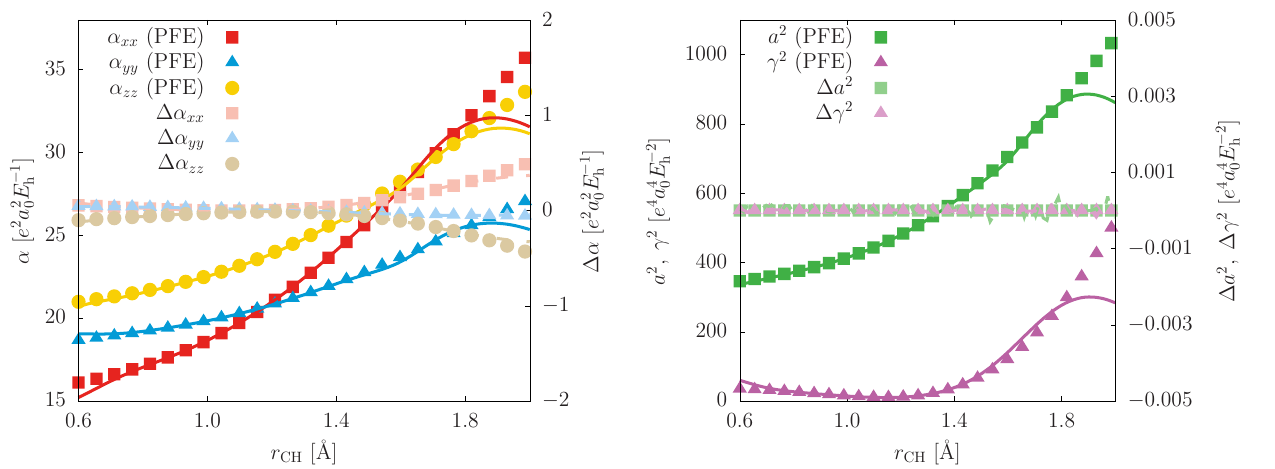}
  \caption{%
    Polarizability of CH$_3$OH vs. $r_4$: components and squared mean (isotropic) and anisotropic polarizabilities, as a function of the $r_4$ \ce{CH} bond distance with all other coordinates fixed at one of the equilibrium structures of PES2025.\cite{Sunaga2025JCP_PES}
    PFE: path-following Eckart frame, 
    $\Delta \alpha = \alpha(\mr{pBF}) - \alpha(\mr{PFE})$
    (and similar for $\Delta a^2$ and $\Delta \gamma^2$), where pBF refers to the primitive body-fixed frame (Sec.~\ref{sec:coord}). 
    The curves show the property surface, the points are the \emph{ab initio} values (at the same CCSD/aug-cc-pVTZ level of theory as the training dataset).
 }\label{fig:r4_CH_pol}
\end{figure*}

Figs.~\ref{fig:r4_CH_mu} and \ref{fig:r4_CH_pol} show the cut of the property surfaces along the $r_4$ CH stretching coordinate (one of the methyl hydrogens not in the $C_s$ symmetry plane, Fig.~\ref{fig:methanol_cord}) while all other coordinates are fixed at their equilibrium value of PES2025. The region near the equilibrium structure can be reproduced by the property surface well for all values of $\mu$ and $\alpha$. 
Of course, the $|\mu|$ length of the vector and the isotropic polarizability, $a$ (proportional to the trace of the polarizability), and $\gamma^2$ (a Placzek invariant) are invariant to the choice of the body-fixed frame. 

For the dipole and polarizability components, we observe a clear frame dependence in the figures, and the vibrational transition integrals and (approximate) vibrational intensities in the later sections exhibit this frame dependence. All further results in this paper are reported for the path-following Eckart frame, Eq.~\eqref{eq:path-following_Eckart}. The rovibrational transition integrals are frame-independent quantities (observables), and they will be computed in future work.

%
%

\subsection{Vibrational energies, vibrational electric dipole and polarizability transition integrals up to the O-H stretching fundamental vibration}
\noindent %
An excerpt of the computed vibrational band origins with assignments and transition dipole and polarizability matrix elements is shown in Table~\ref{tbl:ene_mu2}. In the table, we collect the transitions from the zero-point vibration (ZPV) to the fundamental vibrational states.  (The complete list, including all 675 states, as obtained in the present $b=7$ computation, is provided as \SM.)
The vibrational fundamentals are in excellent agreement with gas-phase experimental data~\cite{Moruzzi1995_ch3oh,Lees2002PRA,Temsamani2003JMS,Lees2004JMS,Xu1997JMS_CH_strech,Hunt1998JMS_OH_stretch,Wang1998JCP_fit}: the root-mean-square deviations between theory and (gas-phase) experiment for the fundamentals and corresponding tunnelling splittings are 2.2~$\mathrm{cm}^{-1}$\ and 0.8~$\mathrm{cm}^{-1}$\, respectively.
We think that the experiment-theory agreement for the higher-energy fundamentals (C-H and O-H stretching range) is too good, most likely due to a fortuitous cancellation of errors. On one hand, PES2025 has various sources of error (discussed in Sec. III.B of Ref.~\citenum{Sunaga2025JCP_PES}), and on the other hand, the limited size of the $b=7$ vibrational basis set can cause several cm$^{-1}$ error in converging the vibrational part of the problem (Fig.~S1 of Ref.~\citenum{Sunaga2024JCTC}). 

Some further comments regarding the $\nu_6$ fundamental vibration and its tunnelling splitting are necessary. We observe strong mixing of the basis functions, \emph{e.g.,} the mixing of $6_1,7_1,11_1$, and torsional functions. The non-degenerate fork of the $\nu_6$ fundamental vibration is assigned to 1319.9~$\mathrm{cm}^{-1}$\ (No.~27~state in Table~S6, Ref.~\citenum{Sunaga2025JCP_PES}), but the $6_1$ basis function has non-negligible contributions to two degenerate pairs within $\pm 25$~$\mathrm{cm}^{-1}$. 
There are degenerate states with strong $6_1$ contribution at 1298.2~$\mathrm{cm}^{-1}$\ (No.~25-26)
and at 1343.5~$\mathrm{cm}^{-1}$\ (No.~29-30), which correspond to --21.1~$\mathrm{cm}^{-1}$\ and +23.6~$\mathrm{cm}^{-1}$\ tunnelling splittings, respectively, depending on which state we assign to the $\nu_6$ tunnelling manifold. Sibert and Castillo-Char\'a \cite{Sibert2005JCP} also noted (in the footnote to their Table~VI) the strong mixing of $\nu_6$. Ref.~\citenum{Bowman2007JPCA}, using the reaction path-Hamiltonian, reported the $\nu_6$ tunnelling splitting to be +20.01~$\mathrm{cm}^{-1}$\ (without further discussion of possible mixing). 

For the dipole transition moments, we list the experimental values taken from TABLE~VIII of Ref.~\citenum{Gribakin2017PRA_methanol}, which compiled data from liquid phase experiments.\cite{Dang-Nhu1990JMS,Florian1997MP,Bertie1997JMS} 
The $\left<\mu_\mr{fi}\right>^2$ matrix element qualitatively reproduces the general trends seen in the data available from the liquid phase, but large deviations (most likely, due to the condensed-phase environment in that experimental dataset) are seen in the $e$-type modes (\emph{e.g.,} \ce{CH3} antisymmetric stretch, $\nu_2$ and $\nu_9$). 

%
%
\begin{table*}
\caption{
All fundamental vibrations of CH$_3$OH: vibrational intervals, $\tilde{\nu}$ in $\mathrm{cm}^{-1}$\, transition dipole integral, $\left<\mu_\mr{fi}\right>^2$ in $e^2a^2_0$,
Raman intensity $A^\mr{R}_\mr{fi}$ in $e^4a^4_0 E^{-2}_\mr{h}$, and
depolarization ratio $\rho_\mr{fi}$,
from the vibrational ground state. 
$\left<\mu_\mr{fi}\right>^2$, $A^\mr{R}_\mr{fi}$, and $\rho_\mr{fi}$ for the $A_1$ and $A_2$ states ($E$ state) are the matrix elements from the vibrational ground state (first excited $E$ state).  
The GENIUSH-Smolyak method with the $b=7$ basis set and PES2025~\cite{Sunaga2025JCP_PES} were used in the vibrational computation. $\left<\mu_\mr{fi}\right>^2$, $A^\mr{R}_\mr{fi}$, and $\rho_\mr{fi}$ correspond to the path-following Eckart frame (Sec.~\ref{sec:pathEckart}). 
Experimental (Exp.) values for $\tilde{\nu}$ and $\left<\mu_\mr{fi}\right>^2$ are from gas- and liquid-phase data, respectively. 
$\Delta$ is the tunneling splitting and $\delta$ labels the deviation from the experimental value. 
The root-mean-square deviations for the vibrational energies and tunneling splitting of the fundamental modes are 2.2 $\mathrm{cm}^{-1}$\ and 0.8 $\mathrm{cm}^{-1}$, respectively.
}
\label{tbl:ene_mu2}
\centering
\begin{tabular}{@{} cc| S[table-format=5.2] c  S[table-format=5.2] c |cc |cc @{}}
\hline\hline
 &  & \multicolumn{4}{c|}{$\tilde{\nu}$} & \multicolumn{2}{c|}{$\left<\mu_\mr{fi}\right>^2$} & $A^\mr{R}_\mr{fi}$ & $\rho_\mr{fi}$ \tabularnewline 
\multicolumn{2}{c|}{State}  & {This Work} &  {$\delta$} & {Exp.} & Ref. & This Work & Exp.~\cite{Gribakin2017PRA_methanol}$^\ast$ & \multicolumn{2}{c}{This Work} \tabularnewline
\hline 
$\nu_{12}$ & $A_2$ & 295.1 & (--0.6) & 294.5 & \citenum{Moruzzi1995_ch3oh} & 9.30[--5] & 2.45{[}--2{]}  & 0.26 & 0.750\tabularnewline
tor(OH) & $E$ & 210.9 & (--2.0) & 208.9 & \citenum{Moruzzi1995_ch3oh} & 6.01[--5] &  & 0.16 & 0.750 \tabularnewline
 & $\Delta$ & -84.2 & (--1.4) & -85.6 &  &  & &&\\[2mm]

$\nu_8$ & $A_1$ & 1034.5 & (--0.1) & 1034.4 & \citenum{Lees2002PRA} & 6.54$[-3]$ & 5.92{[}$-3${]}  &  3.46  &  0.232 \tabularnewline
$\nu(\mathrm{CO})$  & $E$ & 1043.3 & (--0.7) & 1042.6 & \citenum{Lees2002PRA} & 6.57$[-3]$ &  &  3.02  &  0.232 \tabularnewline
 & $\Delta$ & 8.8 & (--0.6) & 8.2 &  &  & \\[2mm]
$\nu_7$ & $A_1$ & 1074.7 & (0.0) & 1074.7 & \citenum{Lees2002PRA} & 1.02$[-4]$ & 6.95{[}$-4${]}  &  4.17  &  0.208 \tabularnewline
$\rho\left(\mathrm{CH}_3\right)$ & $E$ & 1079.1 & (0.2) & 1079.3 & \citenum{Lees2002PRA} & 3.31$[-5]$ &   &  4.03  &  0.210 \tabularnewline
 & $\Delta$ & 4.4 & (0.2) & 4.6 &  &  & \\[2mm]
$\nu_{11}$ & $A_2$ & 1162.1 & (1.9) & 1164.0 & \citenum{Lees2002PRA} & 1.77$[-6]$ & 7.62{[}$-5${]}  &  0.64  &  0.750 \tabularnewline
$\rho\left(\mathrm{CH}_3\right)$  & $E$ & 1155.0 & (1.5) & 1156.5 & \citenum{Lees2002PRA} & 1.61$[-6]$ &   &  0.59  &  0.750 \tabularnewline
 & $\Delta$ & -7.1 & (--0.4) & -7.5 &  &  & \\[2mm]
$\nu_6$ & $A_1$ & 1319.9 & (0.7) & 1320.6 & \citenum{Lees2004JMS} & 6.14$[-4]$ & 9.12{[}$-4${]}&  0.35  &  0.395 \tabularnewline
 $\delta(\mathrm{COH})$ & $E$ & 1298.2$^\text{a}$ & (--0.7) & 1297.5 & \citenum{Lees2004JMS} & 2.42$[$--4$]$ & &  0.14  &  0.323 \tabularnewline
 & $\Delta$ & -21.7 & (--1.4) & -23.1 &  &  & \\[2mm]
$\nu_5$ & $A_1$ & 1450.3 & (3.0) & 1453.3 & \citenum{Temsamani2003JMS} & 1.48$[-4]$ & 1.31{[}$-4${]}&  0.44  &  0.206 \tabularnewline
$\delta\left(\mathrm{CH}_3\right)_{\text {sym }}$ & $E$ & 1458.7 & (3.4) & 1462.1 & \citenum{Temsamani2003JMS} & 1.41$[-4]$ & &  0.47  &  0.253 \tabularnewline
 & $\Delta$ & 8.4 & (0.4) & 8.8 &  &  & \\[2mm]
$\nu_{10}$ & $A_2$ & 1477.8 & (3.7) & 1481.5 & \citenum{Temsamani2003JMS} & 1.29$[-4]$ & 2.07{[}$-4${]}&  2.59  &  0.750 \tabularnewline
$\delta\left(\mathrm{CH}_3\right)_{\text {asym }}$ & $E$ & 1471.7 & (2.2) & 1473.9 & \citenum{Temsamani2003JMS} & 9.67$[-5]$ &&  1.92  &  0.750 \tabularnewline
 & $\Delta$ & -6.1 & (--1.5) & -7.6 &  &  & \\[2mm]
$\nu_4$ & $A_1$ & 1485.4 & (0.7) & 1486.1 & \citenum{Temsamani2003JMS} & 1.74$[-4]$ & 3.11{[}$-4${]}&  2.53  &  0.748 \tabularnewline
$\delta\left(\mathrm{CH}_3\right)_{\text {asym }}$ & $E$ & 1481.8 & (1.5) & 1483.3 & \citenum{Temsamani2003JMS} & 1.31$[-4]$ & &  1.85  &  0.746 \tabularnewline
 & $\Delta$ & -3.6 & (0.8) & -2.8 &  &  & \\[2mm]
$\nu_3$ & $A_1$ & 2841.8 & (2.9) & 2844.7 & \citenum{Wang1998JCP_fit} & 4.34$[-4]$ & 5.13{[}$-4${]}&  21.05  &  0.014 \tabularnewline
$\nu\left(\mathrm{CH}_3\right)_{\text {sym }}$  & $E$ & 2850.7 & (3.1) & 2853.8 & \citenum{Wang1998JCP_fit} & 2.27$[-4]$ & &  10.92  &  0.016 \tabularnewline
 & $\Delta$ & 8.9 & (0.2) & 9.1 &  &  & \\[2mm]
$\nu_9$ & $A_2$ & 2964.2 & (2.5) & 2966.7 & \citenum{Wang1998JCP_fit} & 3.70$[-4]$ & 8.21{[}$-4${]}&  4.34  &  0.750 \tabularnewline
$\nu\left(\mathrm{CH}_3\right)_{\text {asym }}$ & $E$ & 2959.1 & (2.1) & 2961.2 & \citenum{Wang1998JCP_fit} & 1.99$[-4]$ &&  3.41  &  0.117 \tabularnewline
 & $\Delta$ & -5.1 & (--0.4) & -5.5 &  &  & \\[2mm]
$\nu_2$ & $A_1$ & 3002.9 & (4.1) & 3007.0 & \citenum{Xu1997JMS_CH_strech} & 4.41$[-4]$ & 5.66{[}$-4${]}&  10.14  &  0.428 \tabularnewline
$\nu\left(\mathrm{CH}_3\right)_{\text {asym }}$  & $E$ & 2999.8 & (3.9) & 3003.7 & \citenum{Xu1997JMS_CH_strech} & 3.60$[-4]$ &   &  10.97  &  0.346 \tabularnewline
 & $\Delta$ & -3.1 & (--0.2) & -3.3 &  &  & \\[2mm]
$\nu_1$ & $A_1$ & 3686.4 & (--1.1) & 3685.3 & \citenum{Hunt1998JMS_OH_stretch} & 4.01$[-4]$ & 4.77{[}$-4${]}&  15.17  &  0.155 \tabularnewline
$\nu(\mathrm{OH})$  & $E$ & 3692.6 & (--1.0) & 3691.6 & \citenum{Hunt1998JMS_OH_stretch} & 3.92$[-4]$ & &  15.18  &  0.159 \tabularnewline
 & $\Delta$ & 6.2 & (0.1) & 6.3 &  &  & \tabularnewline
 \hline\hline
\end{tabular}
\\[1mm]
$^\ast$ Ref.~\citenum{Gribakin2017PRA_methanol} compiled data from the experimental sources, Refs.~\citenum{Dang-Nhu1990JMS,Florian1997MP,Bertie1997JMS}. \\
$^\text{a}$ We note that the $E$ state at 1343.5~$\mathrm{cm}^{-1}$\ (No.~29-30, Table~S7 of Ref.~\citenum{Sunaga2025JCP_PES}) also has strong $6_1$ contribution and could be assigned to the $\nu_6$ fundamental (see also text and Table~S7 of Ref.~\citenum{Sunaga2025JCP_PES}).
\end{table*}

\begin{figure*}[htbp!]
  \centering
  \includegraphics[width=1.0\linewidth]{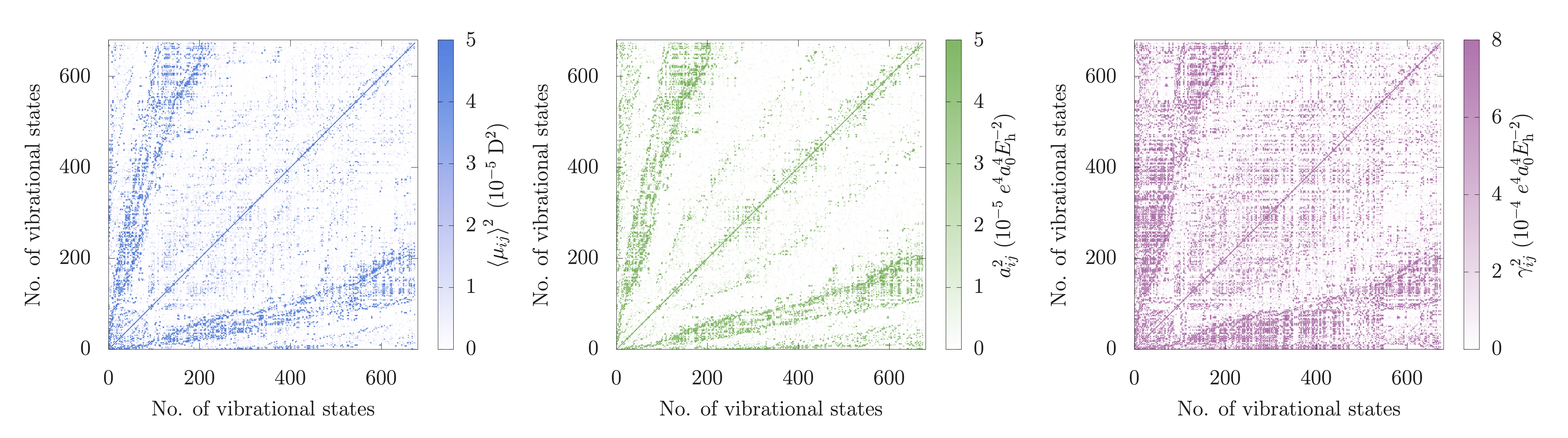}
  \caption{%
    Visual representation of all computed vibrational transition matrix elements for $\left<\mu\right>^2$, $a^2$, and $\gamma^2$ including all the 675 vibrational states computed up to the O-H stretching fundamental region. 
    The GENIUSH-Smolyak method with the $b=7$ basis set \cite{Sunaga2024JCTC} and the \emph{ab initio} PES, PES2025 of Ref.~\citenum{Sunaga2025JCP_PES}, are employed for the vibrational computations. 
  }
    \label{fig:color_plot}
\end{figure*}

All transition matrix elements, for both the electric dipole and the polarizability, are computed with the 675 vibrational states obtained in this work. First, the transition integrals, $\left< \mu_a\right>$ and $\left< \alpha_{ab}\right>$, are evaluated in the path-following Eckart frame; further quantities, \emph{e.g.,} $\left<\mu\right>^2$, $a^2$, and $\gamma^2$ are constructed from these integrals. Figure~\ref{fig:color_plot} highlights the computed data for all transitions of the 675 states; the full dataset (as plain text) is provided as \SM. 

In addition to the transitions from the ZPV to vibrations with significant contributions to the fundamental modes, we can observe non-negligible matrix elements connecting fundamentals, overtones, and combination bands. 
Large matrix elements are obtained for 
(a) the torsional excitation and 
(b) mode coupling states, \emph{e.g.,} %
(a) No. 10 $(E,\nu_\tau=3,0)$ and No. 599 $(E,\nu_\tau=3,3_1,5_2)$ states, which correspond to the excitation of the $3_1$ mode, and 
(b) No. 20 $(E,\nu_\tau=0,7_1,11_1)$ and No. 190 $(E,\nu_\tau=0,7_1+5_1)$ states, which correspond to the excitation of the $5_1$ mode, while other modes are also changed. 
The coloured plots for $\left<\mu\right>^2$ and $a^2$ look similar at first glance, but the isotropic term $a^2$ becomes small when some of the $\left<\alpha_{ii}\right>$ components have opposite signs with similar absolute values (see \Eq{eq:a_gamma}). Interestingly, the figure for $\gamma^2$ shows more matrix elements with relatively large values, although its absolute values for a specific configuration are smaller than those of $a^2$ (\FIG{fig:1D_tor_pol} and \ref{fig:r4_CH_pol}).

%
%

\subsection{Vibrational infrared and Raman spectra \label{sec:spectra}}
\begin{figure}[htbp!]
    \centering
    \includegraphics[width=0.4\linewidth]{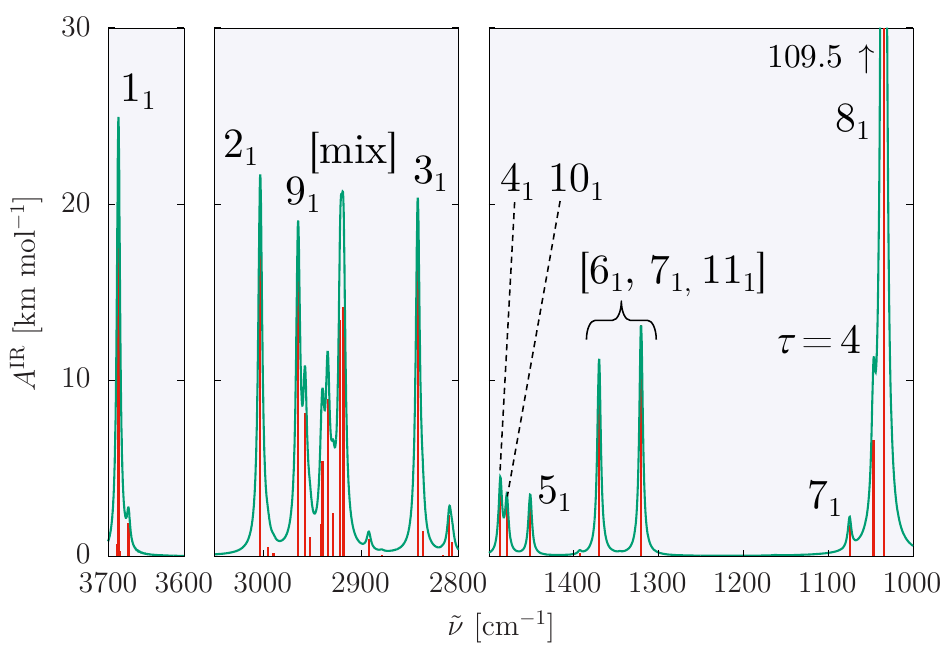}
    \caption{%
    Vibrational infrared spectrum of methanol with transitions from the vibrational ground state. ``$[\ldots]$'' labels strongly mixed states. The vibrational intensities are computed from the electric dipole moment in the path-following Eckart frame. The GENIUSH-Smolyak method with $b=7$ and the PES of Ref.~\citenum{Sunaga2025JCP_PES} are used in the vibrational computations.
    The stick spectrum (in red), calculated according to Eq.~\eqref{eq:A_IR} is convoluted (in green) with a Lorentz distribution of 5~$\mathrm{cm}^{-1}$\ full width at half maximum (FWHM).
    }
    \label{fig:IR_spectra}
\end{figure}

\begin{figure*}[htbp!]
    \centering
    \includegraphics[width=0.8\linewidth]{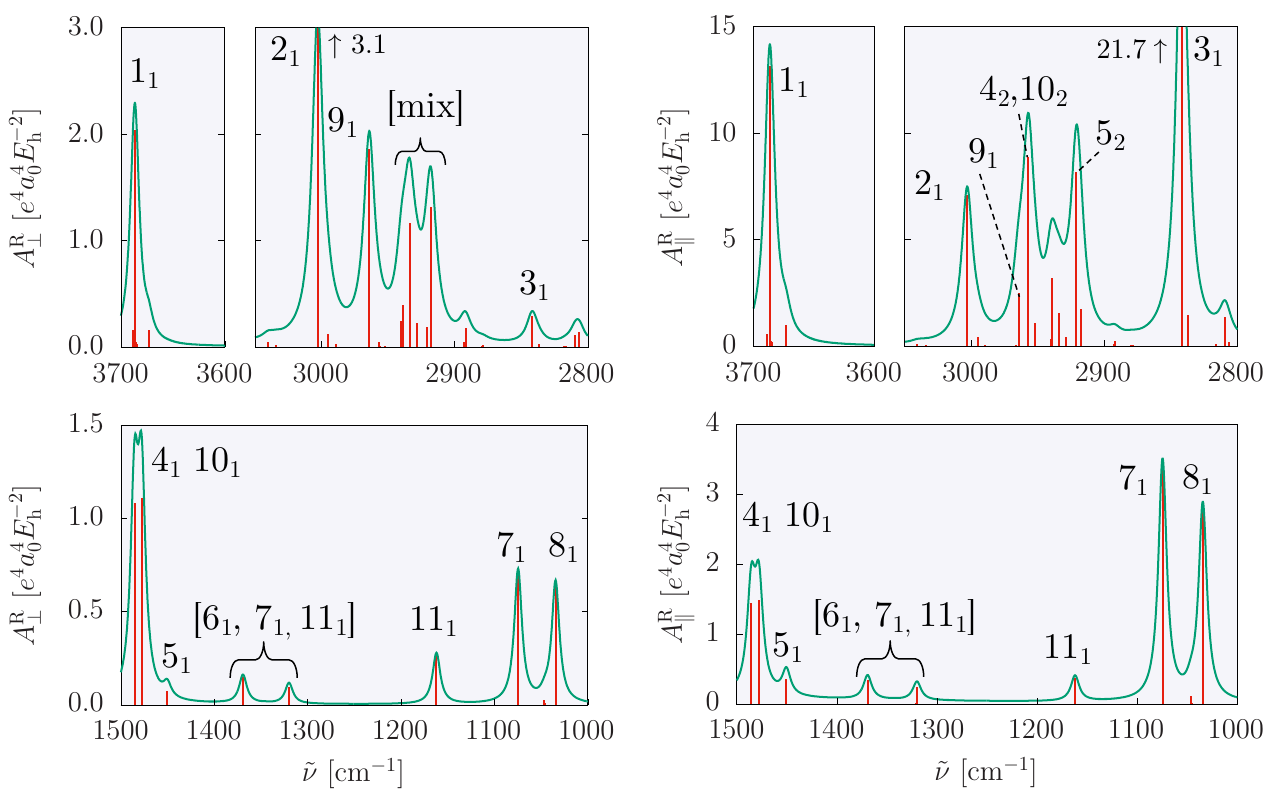}
    \caption{Vibrational Raman spectrum of methanol with transitions from the vibrational ground state. ``$[\ldots]$'' labels strongly mixed states. The vibrational intensities were computed from body-fixed polarizabilities corresponding to the path-following Eckart frame. The GENIUSH-Smolyak method with $b=7$ and the PES of Ref.~\citenum{Sunaga2025JCP_PES} were used in the vibrational computations. The stick spectrum (in red), calculated according to Eq.~\eqref{eq:A_Raman} is convoluted (in green) with a Lorentz distribution 
    of 10~$\mathrm{cm}^{-1}$\ full width at half maximum (FWHM).}
    \label{fig:Raman_spectra}
\end{figure*}

Figures~\ref{fig:IR_spectra} and~\ref{fig:Raman_spectra} show the simulated infrared (IR) and Raman spectra of the methanol molecule, respectively, with transitions from the vibrational ground state. The simulated spectra can correspond to low-temperature, vibrational infrared jet-spectroscopy or matrix isolation observations (\emph{e.g.,} methanol monomer~\cite{Perchard2007CP,Perchard2008CP_2400,Kollipost2014JCP_mono_dim,Jiang2017RSC_complex,Dinu2024ACSAu} or clusters~\cite{Kollipost2014JCP_mono_dim,Heger2016PCCP_dimer,Jiang2017RSC_complex,Schweer2023PCCP_FA_methanol}). Lorentz profiles with
5~$\mathrm{cm}^{-1}$~and 10~$\mathrm{cm}^{-1}$~of the full-width-at-half-maximum (FWHM) are employed for the IR and Raman spectral plots (Figs.~\ref{fig:IR_spectra} and \ref{fig:Raman_spectra}), respectively. 

Strong peaks are assigned following the traditional vibrational labelling (corresponding to the \Cs\ point group), but several differences are found in our vibration-torsion computation based on the \Ctv\ molecular symmetry group. Several $a'$ and $a''$ type motions of \Cs\ (e.g., asymmetric stretch of \ce{CH3}, $2_1$ and $9_1$) belong to degenerate $e$-type motion in \Ctv, and they are mixed in our vibrational wavefunctions. 
In the figures, we observe several band systems with contribution from fundamentals. 
In the 1050-1400 $\mathrm{cm}^{-1}$~range, the torsional ground and excited states, and small-amplitude motions are strongly mixed. 

Strong peaks in the combination band and overtone regions are observed due to mixing among the fundamental, overtone and combination modes. For example, the strong mixing with $3_1$ is found with $5_2$ and $4_2,10_2$, which is the reason for the strong peaks of these states in the parallel Raman spectrum, $A^{\mr{R}}_\parallel$. 
The peak $11_1$ at around 1162~$\mathrm{cm}^{-1}$\ is very weak and invisible in the scale of the IR spectrum,  \FIG{fig:IR_spectra},
but it is visible in the Raman spectrum. 
The details of the mode-coupling features are provided in the Tables S6--S11 of the \SM. 

%
%

%
%

\section{Summary and conclusion}\label{sec:conclusion}
\noindent%
This work reported the development of the electric dipole moment and polarizability surfaces of the methanol molecule using an \textit{ab initio} dataset computed at the CCSD/aug-cc-pVTZ level of theory and equivariant neural networks as available in the MACE program.\cite{Batatia2022_proc_MACE} The equivariant representation\cite{Bartok2010arXiv_rot_invariant,Bartok2010PRL_GAP} ensures the exact representation of both the permutational symmetry of identical nuclei and rotational covariance of the tensorial quantities.
As the first application of the computed property surfaces, the low-temperature vibrational infrared and Raman spectra were determined up to the OH-stretching region of methanol.
For the spectral intensity computations, the vibrational energies and wave functions were obtained in continued variational vibrational computations with the GENIUSH-Smolyak program \cite{Avila2019JCP_methodology,Avila2019JCP_CH4F-} (using the $b=7$ basis set of Ref.~\citenum{Sunaga2024JCTC}) and PES2025.\cite{Sunaga2025JCP_PES} 
Despite the accuracy limitations of PES2025 (Sec.~III.B of Ref.~\citenum{Sunaga2025JCP_PES}) and the $b=7$ vibrational basis (allowing a maximal 7 quantum excitation in the harmonic oscillator basis describing the small-amplitude vibrations), the vibrational fundamentals and tunneling splittings are obtained in excellent agreement with gas-phase experimental data, with 2.2~$\mathrm{cm}^{-1}$\ and 0.8~$\mathrm{cm}^{-1}$\ root-mean-square error, respectively.
The developed property surfaces enable direct comparison of future rovibrational computations with high-resolution spectroscopy data, and represent an important step towards developing a line list \cite{Exomol1,Exomol2} for methanol for helping to explore the chemical and physical environment of the universe.

\section{Acknowledgement}
\noindent%
A.S. thanks the European Union's Horizon 2022 Research and Innovation Programme under the Marie Skłodowska-Curie Grant Agreement No.~101105452. We also thank the Hungarian National Research, Development, and Innovation Office (FK~142869 and NKFI~153229) for financial support. We acknowledge DKF (Governmental Agency for IT Development, Hungary) for awarding us access to the Komondor HPC facility based in Hungary.

%

\clearpage
\input{arXiv_som_inc}

\end{document}

%% file: arXiv_som_inc.tex
\clearpage
\begin{center}
{\large
\textbf{Supplemental Material}
}\\[0.25cm]
{\large
\textbf{Vibrational infrared and Raman spectra of the methanol molecule with equivariant
neural-network property surfaces}
} \\[0.5cm]

Ayaki Sunaga,$^1$ Albert P. Bartók,$^{2,3}$ Edit Mátyus$^{1,\ast}$ \\[0.25cm]
\emph{$^1$~ELTE, E\"otv\"os Lor\'and University, Institute of Chemistry, P\'azm\'any P\'eter s\'et\'any 1/A 1117 Budapest, Hungary} \\[0.25cm]
\emph{$^2$~Department of Physics, University of Warwick, Coventry, CV4 7AL, UK} \\[0.25cm]
\emph{$^3$~Warwick Centre for Predictive Modelling, School of Engineering, University of Warwick, Coventry, CV4 7AL, UK} \\[0.25cm]

$^\ast$edit.matyus@ttk.elte.hu 

~\\[0.15cm]
(Dated: 30 April 2026)
\end{center}

\setcounter{section}{0}
\renewcommand{\thesection}{S\arabic{section}}
\setcounter{subsection}{0}
\renewcommand{\thesubsection}{S\arabic{section}.\arabic{subsection}}

\setcounter{equation}{0}
\renewcommand{\theequation}{S\arabic{equation}}

\setcounter{table}{0}
\renewcommand{\thetable}{S\arabic{table}}

\setcounter{figure}{0}
\renewcommand{\thefigure}{S\arabic{figure}}

\clearpage
%
%
\section{Development process of the property surfaces}
To obtain the training sets for fitting and test sets for evaluating the surfaces' quality, we first carry out the electronic structure computation for the 39~401 geometries in the fitting set obtained in Ref.~\citenum{Sunaga2025JCP_PES}. The property surface of the $\alpha_{zz}$ component was unstable when we performed the fit using all the geometry points. Therefore, we extracted the nine geometries that showed negative eigenvalues during the linear response computation of polarizability; these points corresponded to largely distorted structures, and the $|\alpha_{zz}|$ values exceeded 90~$e^2a^2_0E^{-1}_\mr{h}$. In the end, we included 35~000 geometries in the training set, and the remaining 4~392 geometries formed the test set. These geometries were randomly extracted from the total 39~392 geometries, and three randomly extracted sets were tested. The qualities of these surfaces are summarized in Tables~\ref{tbl:RMSE_SM} and~\ref{tbl:MAPE_SM}. The key input parameters of MACE employed in the fitting are provided in Table~1 of the main text. 

Based on comparison of the developed property surfaces, SET1 is employed for the dipole moment surface, and SET3 is employed for the polarizability surface in the computation shown in the main text and in the following sections. However, no significant difference is observed in the RMSE errors of the different SETs. This observation suggests that the results are insensitive to precise manner the geometry points are extracted from the geometry set of 39~401 points. The accuracy of the dipole moment and polarizability surfaces obtained with SET1-3 is sufficient and comparable to that of the quantum-chemical computations. 

The dependence of the training results on hyperparameters is provided in Tables~\ref{tbl:RMSE_param} and \ref{tbl:MAPE_param}. 
Although slight improvement (\emph{i.e.,} the decrease of the RMSE and MAPE) is observed in the polarizability surface trained with $B=64$ and $\nu=4$ compared with those listed in Tables~\ref{tbl:RMSE_SM} and~\ref{tbl:MAPE_SM}, the parameter dependence is much smaller than the diagonal components at the equilibrium, ca. 20~$e^2a^2_0E^{-1}_\mr{h}$ (Table~2 of the main text). For the dipole moment surfaces, the parameters employed in (SET1) are the best choice among the parameters investigated in Tables~\ref{tbl:RMSE_SM} and~\ref{tbl:MAPE_SM}.

The MACE program~\cite{Batatia2022_proc_MACE} at commit hash a8f2f6cd64e7 was used for the fitting of the property surfaces.

%
%
\begin{table}[htbp!]
\caption{
Root mean square error (RMSE) of the polarizability and dipole moment surfaces for methanol in $e^2a^2_0E^{-1}_\mr{h}$ for polarizability and $ea_0$ for dipole moment, fitted by the MACE program.
The \textit{ab} \textit{initio} values are obtained at the CCSD/aug-cc-pVTZ level of theory.
}\label{tbl:RMSE_SM}
\begin{tabular}{@{} ccccccccccc @{}}
\hline\hline
 &  & \multicolumn{6}{c}{polarizability} & \multicolumn{3}{c}{dipole moment}\tabularnewline
 &  & $xx$ & $yy$ & $zz$ & $xy$ & $yz$ & $xz$ & $x$ & $y$ & $z$\tabularnewline
 \hline
SET1 & training & 0.047 & 0.053 & 0.087 & 0.029 & 0.048 & 0.032 & 0.0018 & 0.0028 & 0.0028\tabularnewline
 & test & 0.049 & 0.056 & 0.103 & 0.028 & 0.048 & 0.029 & 0.0018 & 0.0026 & 0.0031\tabularnewline
SET2 & training & 0.048 & 0.054 & 0.082 & 0.029 & 0.046 & 0.032 & 0.0020 & 0.0030 & 0.0030\tabularnewline
 & test & 0.044 & 0.048 & 0.080 & 0.031 & 0.044 & 0.029 & 0.0018 & 0.0035 & 0.0030\tabularnewline
SET3 & training & 0.048 & 0.052 & 0.079 & 0.029 & 0.046 & 0.032 & 0.0019 & 0.0028 & 0.0029\tabularnewline
 & test & 0.049 & 0.057 & 0.080 & 0.028 & 0.046 & 0.030 & 0.0017 & 0.0026 & 0.0029\tabularnewline
 \hline\hline
\end{tabular}
\end{table}

\begin{table}[htbp!]
\caption{
Mean absolute percentage error (MAPE,~\%) of the polarizability and dipole moment surfaces for methanol, fitted by MACE program. The \textit{ab} \textit{initio} values are obtained at the CCSD/aug-cc-pVTZ level of theory.
}\label{tbl:MAPE_SM}
\begin{tabular}{@{} ccc ccc ccc cc @{}}
\hline\hline
 &  & \multicolumn{6}{c}{polarizability} & \multicolumn{3}{c}{dipole moment}\tabularnewline
 &  & $xx$ & $yy$ & $zz$ & $xy$ & $yz$ & $xz$ & $x$ & $y$ & $z$\tabularnewline
 \hline
SET1 & training & 0.15 & 0.16 & 0.16 & 21.82 & 19.97 & 17.22 & 42.95 & 0.41 & 1.34\tabularnewline
 & test & 0.15 & 0.16 & 0.16 & 53.84 & 17.50 & 20.63 & 20.86 & 0.40 & 1.05\tabularnewline
SET2 & training & 0.15 & 0.16 & 0.16 & 20.54 & 17.12 & 18.81 & 54.47 & 0.42 & 1.58\tabularnewline
 & test & 0.15 & 0.16 & 0.16 & 41.98 & 14.52 & 21.66 & 24.73 & 0.48 & 1.00\tabularnewline
SET3 & training & 0.15 & 0.16 & 0.16 & 27.11 & 19.06 & 18.68 & 26.48 & 0.41 & 1.24\tabularnewline
 & test & 0.16 & 0.16 & 0.16 & 19.79 & 29.20 & 16.00 & 175.37 & 0.42 & 1.19\tabularnewline
 \hline\hline
\end{tabular}
\end{table}

%
%
\begin{table}
\caption{
Root mean square error (RMSE) of the polarizability, in $e^2a^2_0E^{-1}_\mr{h}$, and dipole moment, in $ea_0$, fitted with the MACE program using various hyperparameters. $B$: batch size, $\nu$: correlation order. All other parameters are the same as in Table~1 of the main text.
The \textit{ab} \textit{initio} values are obtained at the CCSD/aug-cc-pVTZ level of theory.
}\label{tbl:RMSE_param}
\begin{tabular}{@{} lcccccccccc @{}}
\hline\hline
 &  & $xx$ & $yy$ & $zz $&$ xy $& $yz$ & $xz$ &$ x$ & $y$ &$ z$\tabularnewline
 \hline
$B=64$ & training & 0.055 & 0.060 & 0.087 & 0.033 & 0.033 & 0.037 & 0.0024 & 0.0035 & 0.0034\tabularnewline
 & test & 0.055 & 0.064 & 0.084 & 0.031 & 0.031 & 0.032 & 0.0023 & 0.0033 & 0.0036\tabularnewline
$B=16$ & training & 0.044 & 0.048 & 0.075 & 0.027 & 0.027 & 0.029 & 0.0021 & 0.0029 & 0.0029\tabularnewline
 & test & 0.044 & 0.052 & 0.077 & 0.027 & 0.027 & 0.028 & 0.0021 & 0.0028 & 0.0030\tabularnewline
$\nu=4$ & training & 0.039 & 0.046 & 0.067 & 0.025 & 0.025 & 0.030 & 0.0020 & 0.0029 & 0.0028\tabularnewline
 & test & 0.046 & 0.055 & 0.063 & 0.024 & 0.024 & 0.027 & 0.0019 & 0.0028 & 0.0028\tabularnewline
 \hline\hline
\end{tabular}
\end{table}

\begin{table}
\caption{
Mean absolute percentage error (MAPE,~\%) of the polarizability and dipole moment surfaces for methanol, fitted with the MACE program using various hyperparameters. $B$: batch size, $\nu$: correlation order. All other parameters are the same as in Table~1 of the main text.
The \textit{ab} \textit{initio} values are obtained at the CCSD/aug-cc-pVTZ level of theory.
}\label{tbl:MAPE_param}
\begin{tabular}{@{} lcccccccccc @{}}
\hline\hline
 &  & $xx $&$ yy $&$ zz $& $xy $&$ yz$ & $xz $& $x $& $y $&$ z$\tabularnewline
 \hline
$B=64$& training & 0.18 & 0.18 & 0.17 & 28.40 & 28.40 & 21.01 & 91.97 & 0.52 & 1.58\tabularnewline
 & test & 0.18 & 0.18 & 0.17 & 20.63 & 20.63 & 20.92 & 30.62 & 0.54 & 1.06\tabularnewline
$B=16$& training & 0.14 & 0.14 & 0.15 & 19.26 & 19.26 & 16.59 & 72.35 & 0.44 & 1.34\tabularnewline
 & test & 0.14 & 0.15 & 0.15 & 19.14 & 19.14 & 19.03 & 26.59 & 0.46 & 0.91\tabularnewline
$\nu=4$ & training & 0.12 & 0.14 & 0.14 & 22.62 & 22.62 & 18.00 & 73.14 & 0.43 & 1.31\tabularnewline
 & test & 0.13 & 0.15 & 0.14 & 26.36 & 26.36 & 17.35 & 23.80 & 0.43 & 0.95\tabularnewline
 \hline\hline
\end{tabular}
\end{table}

%
%
\section{Convergence test with the number of geometry points in the training set}

\begin{figure}[htbp!]
    \centering
    \includegraphics[width=0.9\linewidth]{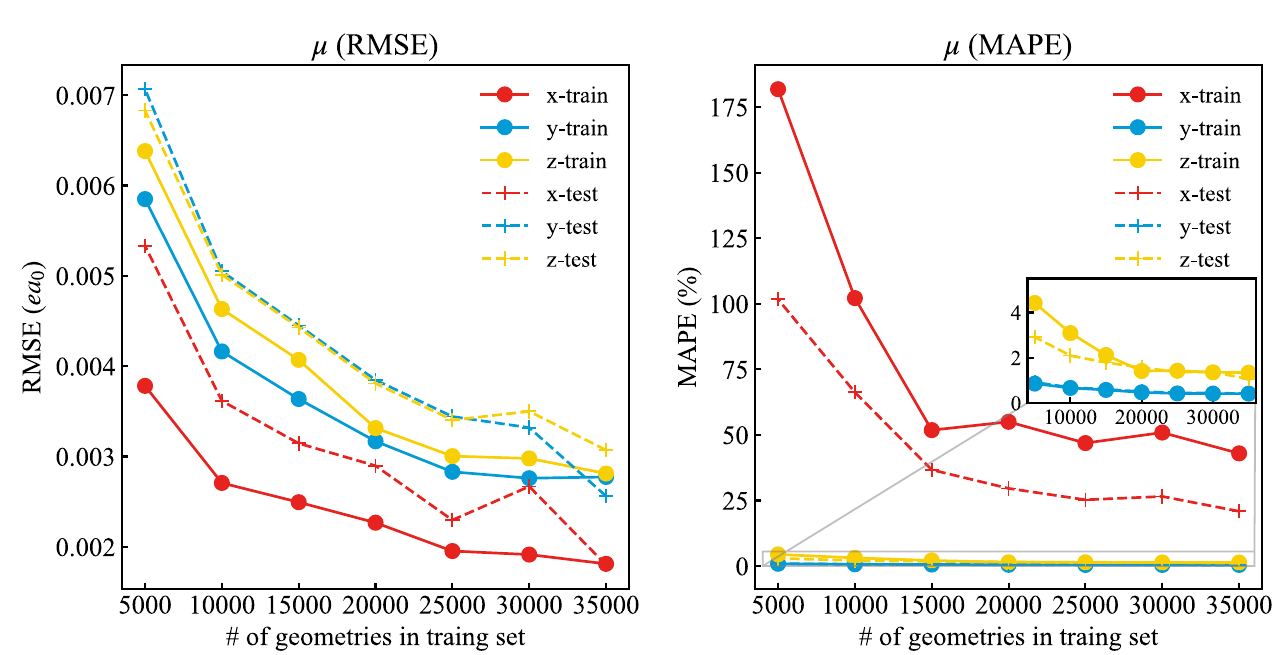}
    \caption{RMSE in $ea_0$ and MAPE in \% for the training set and test set obtained with dipole moment surface of the methanol molecule. The data points were computed at the CCSD/aug-cc-pVTZ level of theory.}
    \label{fig:dip_decrease}
\end{figure}

\begin{figure}[htbp!]
    \centering
    \includegraphics[width=0.9\linewidth]{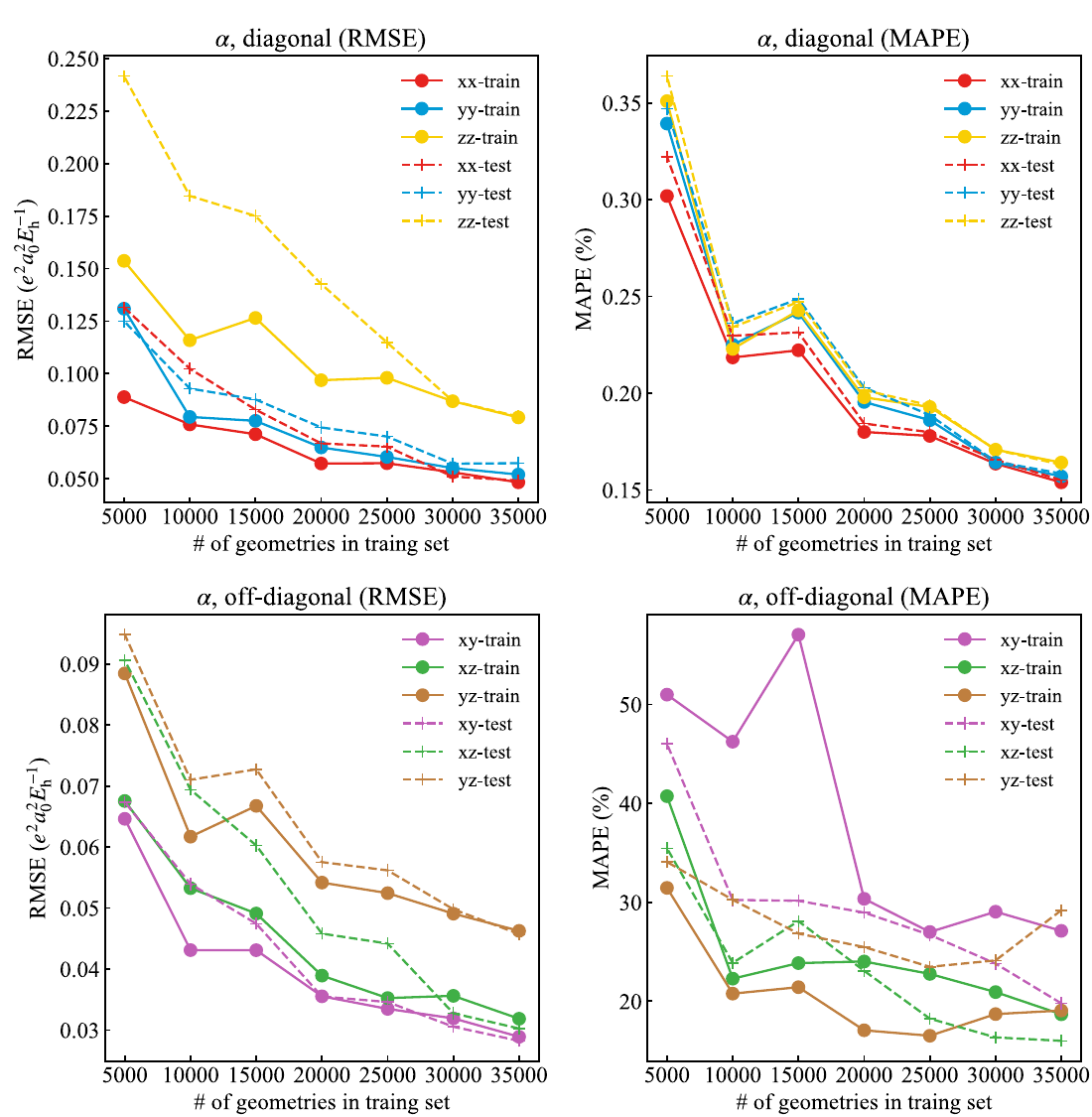}
    \caption{RMSE in $e^2a^2_0 E^{-1}_\mr{h}$ and MAPE in \% for the training set and test set obtained with polarizability surface for the methanol molecule. The data points were computed at the CCSD/aug-cc-pVTZ level of theory.}
    \label{fig:pol_decrease}
\end{figure}

The convergence with respect to the number of geometry points was investigated using SET1 for dipole moment and SET3 for polarizability. We designed the training set with fewer geometries to be a subset of the training set with more geometries. For example, the training set with 30~000 geometries (geometries in the test set) was obtained by randomly extracting 5~000 geometries from the 35~000 geometries in the training set and moving them to the test set (FIGs. \ref{fig:dip_decrease} and \ref{fig:pol_decrease}). 

The RMSE basically decreases as the number of geometry points in the training set increases. The error is not monotonically decreasing due to largely distorted geometries, picked randomly in the training/test sets.  
Most components are converged with respect to the number of geometries, but the RMSE of the off-diagonal components of \pol\ could be decreased further. However, since the RMSE errors of the off-diagonal part are sufficiently smaller than the diagonal errors, we do not have to increase the number of geometries. We should note that the components with large values (the $y$ and $z$ components of $\mu$ and the diagonal components of \pol) are converged with a smaller number of geometry points:  approximately 20~000 geometries (5000 geometries) are sufficient to achieve the error below 2\% (1\%) of the $\mu$ ($\alpha_{ii}$).
The components with small values (the $x$ component of $\mu$ and the off-diagonal components of \pol) show relatively large MAPEs because some of their values are close to zero.

%
%

%
%

\clearpage

%
%
\section{Assignment of the vibrational states and comparison with experiment}

In Tables~\ref{tbl:VBO_I}-\ref{tbl:VBO_6}, we list the vibrational states up to the OH stretching fundamental vibrational as obtained with the GENIUSH-Smolyak method ($b=7$ vibrational basis\cite{Sunaga2024JCTC} and PES2025\cite{Sunaga2025JCP_PES}) and available data originating from gas-phase experiment. 
The first 150 vibrational states were already listed in the Tables~S7--S9 of Ref.~\citenum{Sunaga2025JCP_PES}. We extend it with the further 525 states obtained since the publication of that paper. 
Table~\ref{tbl:harm_freq_aveL} lists the harmonic frequencies of PES2025~\cite{Sunaga2025JCP_PES} and corresponding labels, which allows us to define the vibrational labels used to assign the variational vibrational states.

The vibrational energies of the fundamental modes are listed in Table~3 of the main text, but we note that strong torsion-vibration mixing is observed for $\nu_6$ (No.~25-27), whereas mode coupling is observed for $\nu_7$ (No.~17-19), $\nu_3$ (No.~265, 268-269), and $\nu_2$ (No.~311-313). 
The importance of coupling of $3_1$ and $5_2$ (No.~265), and $9_1$ and $4_1+10_1$ (No.~313) was already pointed out in a vibrational configurational interaction computation (Table~S3 of Ref.~\citenum{Dinu2024ACSAu}). The coupling between the torsional and \ce{CH} stretch motion was also investigated in Refs.~\citenum{Wang1998JCP_fit} and \citenum{Clasp2006JCP_CH}.

%
%
\begin{table}[!htbp]
\caption{%
Harmonic frequencies corresponding to PES2025\cite{Sunaga2025JCP_PES} used in this work. $\Gamma$ is the symmetry label (deliberately in small letters referring to `single mode') corresponding to the \Ctv\ molecular symmetry group.
}
\label{tbl:harm_freq_aveL}
\begin{tabular}{@{}clc cc @{}}
\hline\hline
Label & Description & coord & $\Gamma$ & {$\tilde{\nu}$ [cm$^{-1}$]} \tabularnewline   %
 \hline
$\nu_1$& $\nu$(\ce{OH}) & $r_\mathrm{OH}$ & $a_1$ &   3864.31 \tabularnewline %
$\nu_2$ & $\nu$(\ce{CH3})$_{\mathrm{asym}}$  & $r_\mathrm{CH}$ &$e$ & 3099.06   \tabularnewline %
$\nu_9$  & $\nu$(\ce{CH3})$_{\mathrm{asym}}$ & $r_\mathrm{CH}$&$e$ &  3099.06   \tabularnewline %
$\nu_3$& $\nu$(\ce{CH3})$_{\mathrm{sym}}$& $r_\mathrm{CH}$ &  $a_1$ &   3024.56  \tabularnewline %
$\nu_4$&$\delta$(\ce{CH3})$_{\mathrm{asym}}$ & $\varphi_2$ &  $e$ &   1517.64 \tabularnewline  %
$\nu_{10}$ & $\delta$(\ce{CH3})$_{\mathrm{asym}}$ & $\varphi_1$ &$e$ &   1517.64  \tabularnewline %
$\nu_5$&$\delta$(\ce{CH3})$_{\mathrm{sym}}$ & $\theta_\mathrm{HCO}$ &  $a_1$ & 1486.35   \tabularnewline  %
$\nu_6$&$\delta$(\ce{COH}) & $\theta_\mathrm{COH}$ &  $a_1$ &  1280.06  \tabularnewline  %
$\nu_{11}$& $\rho$(\ce{CH3})$_{\mathrm{}}$& $\theta_\mathrm{HCO}$ &  $e$ &   1183.72  \tabularnewline%
$\nu_7$  &$\rho$(\ce{CH3})$_{\mathrm{}}$ &$\theta_\mathrm{HCO}$ & $e$ &   1183.72  \tabularnewline%
$\nu_8$& $\nu$(\ce{CO})$_{\mathrm{}}$&  $r_\mathrm{CO}$ & $a_1$ &   1076.05  \tabularnewline %
\hline\hline
\end{tabular}
\end{table}

%
%
\begin{table}
\caption{
Vibrational energies, $\tilde{\nu}_\mr{PES}$ in $\mathrm{cm}^{-1}$, referenced to the zero-point vibrational energy (ZPVE, $11119.58 \mathrm{~cm}^{-1}$) of $\mathrm{CH}_3 \mathrm{OH}$ in 12D computed with the GENIUSH-Smolyak program with the $b=7$ basis set and using PES2025.\cite{Sunaga2025JCP_PES} 
The first 150 vibrational states are listed in Tables S7-S9 of the Supplementary Material of Ref.~\citenum{Sunaga2025JCP_PES}.
Comparison with the vibrational band origins derived from (gas-phase) experiment is shown as $\delta=\tilde{\nu}_{\exp }-\tilde{\nu}_{\mathrm{PES}}$, in $\mathrm{cm}^{-1}$. SAMs: assignment of the curvilinear normal modes $1_n, 2_n, \ldots, 11_n(n=0,1, \ldots)$ (Table~\ref{tbl:harm_freq_aveL}), no excitation ($n=0$) is noted as ``0''. ``$[\ldots]$'' labels the largest contribution(s) from strongly mixed states. $\Gamma: C_{3 \mathrm{v}}(\mathrm{M})$ label according to Table~\ref{tbl:harm_freq_aveL} and the irreducible representations of the torsional part. Further details are in Sec.~S1 of the Supporting Information of Ref.~\citenum{Sunaga2024JCTC}. All degeneracies for the $E$ states are converged better than $0.01 \mathrm{~cm}^{-1}$, except for the pairs listed in the footnotes.}\label{tbl:VBO_I}
\begin{tabular}{@{}ccccc| ccccc@{}}
\hline
\hline
\# & $\nu_{\tau}$ & SAMs & $\Gamma$ & $\tilde{\nu}_\mr{PES}$ & \# & $\nu_{\tau}$ & SAMs & $\Gamma$ & $\tilde{\nu}_\mr{PES}$  \tabularnewline
\hline
151-152 & 0,2 & {[}$11_1+7_1,6_1+7_1${]} & $E$ & 2494.3 & 208 & 1 & $6_2$ & $A_2$ & 2676.0\tabularnewline
153-154 & 4 & {[}$4_1,10_1${]} & $E$ & 2497.5 & 209-210 & 1 & $5_1+8_1$ & $E$ & 2690.5\tabularnewline
155 & 3 & $5_1$ & $A_2$ & 2497.9 & 211-212 & 2 & $7_1+8_1,11_1+8_1$ & $E$ & 2694.2\tabularnewline
156 & 4 & $5_1$ & $A_1$ & 2498.7 & 213-214 & 2 & {[}$7_2,6_1+7_1${]} & $E$ & 2695.7\tabularnewline
157-158 & 0 & $4_1+8_1,10_1+8_1$ & $E$ & 2505.1 & 215 & 1 & $10_1+8_1,4_1+8_1$ & $A_2$ & 2706.2\tabularnewline
159-160 & 2 & $7_1+8_1,11_1+8_1$ & $E$ & 2506.7 & 216 & 1 & $4_1+8_1,10_1+8_1$ & $A_1$ & 2715.7\tabularnewline
161 & 0 & $4_1+8_1,10_1+8_1$ & $A_2$ & 2510.5 & 217-218 & 1 & {[}$4_1+8_1,10_1+8_1${]} & $E$ & 2717.8\tabularnewline
162-163 & 0 & $4_1+8_1,10_1+8_1$ & $E$ & 2513.6 & 219 & 2 & $6_2$ & $A_1$ & 2726.4\tabularnewline
164 & 0 & $10_1+8_1,4_1+8_1$ & $A_1$ & 2516.9 & 220-221 & 1,2 & {[}$6_2,11_1+7_1${]} & $E$ & 2727.9\tabularnewline
165 & 0 & {[}$5_1+6_1,5_1+11_1${]} & $A_1$ & 2520.1 & 222-223 & 6 & 0 & $E$ & 2733.6\tabularnewline
166-167 & 0 & {[}$5_1+11_1,5_1+6_1${]} & $E$ & 2523.5 & 224-225 & 0 & $5_1+6_1$ & $E$ & 2744.5\tabularnewline
168-169 & 3,4 & $4_1,10_1$ & $E$ & 2532.1 & 226 & 2 & {[}$7_1+8_1,11_1+8_1${]} & $A_2$ & 2745.0\tabularnewline
170-171 & 0 & {[}$10_1+7_1,10_1+6_1${]} & $E$ & 2542.5 & 227 & 2 & {[}$6_1+11_1,6_1+7_1${]} & $A_1$ & 2759.9\tabularnewline
172 & 0 & {[}$10_1+7_1,4_1+11_1${]} & $A_2$ & 2545.4 & 228 & 1 & {[}$5_1+8_1,5_1+11_1${]} & $A_2$ & 2760.8\tabularnewline
173-174 & 0 & {[}$7_2,4_1+7_1${]} & $E$ & 2548.0 & 229 & 0,1 & {[}$5_1+6_1,5_1+11_1${]} & $A_1$ & 2764.8\tabularnewline
175 & 0 & {[}$4_1+7_1,10_1+11_1${]} & $A_1$ & 2553.0 & 230-231 & 0 & {[}$10_1+6_1,4_1+6_1${]} & $E$ & 2765.8\tabularnewline
176-177 & 1 & {[}$11_1+7_1,11_2${]} & $E$ & 2554.0 & 232 & 0 & $4_1+6_1,10_1+6_1$ & $A_1$ & 2767.2\tabularnewline
178 & 2 & {[}$6_1+8_1,7_1+8_1${]} & $A_1$ & 2562.5 & 233 & 1 & $5_1+8_1$ & $A_2$ & 2769.3\tabularnewline
179-180 & 4 & $11_1,7_1$ & $E$\footnotemark[1] & 2566.2 & 234-235 & 2 & $6_1+8_1$ & $E$ & 2773.8\tabularnewline
181 & 4 & $11_1,7_1$ & $A_2$ & 2567.6 & 236 & 2 & {[}$6_1+8_1,11_1+8_1${]} & $A_1$ & 2776.5\tabularnewline
182 & 1 & $6_1+8_1$ & $A_2$ & 2571.3 & 237 & 1 & {[}$6_1+7_1,6_1+11_1${]} & $A_2$ & 2779.7\tabularnewline
183-184 & 2 & $8_2$ & $E$ & 2574.3 & 238-239 & 0,1 & {[}$4_1+6_1,4_1+7_1${]} & $E$ & 2780.8\tabularnewline
185 & 4 & $11_1,7_1$ & $A_1$ & 2575.4 & 240-241 & 1 & $10_1+8_1,5_1+6_1$ & $E$ & 2788.9\tabularnewline
186 & 4 & {[}$7_1,11_1${]} & $A_2$ & 2585.9 & 242 & 1 & $5_1+8_1$ & $A_2$ & 2790.5\tabularnewline
187 & 2 & $6_1+8_1$ & $A_1$ & 2588.7 & 243-244 & 1 & {[}$4_1+11_1,5_1+6_1${]} & $E$ & 2793.9\tabularnewline
188-189 & 1,2 & {[}$6_2,6_1+7_1${]} & $E$ & 2589.9 & 245-246 & 3 & $8_2$ & $E$ & 2800.7\tabularnewline
190-191 & 0 & $5_1+7_1$ & $E$ & 2597.5 & 247 & 1 & $10_1+6_1,4_1+6_1$ & $A_2$ & 2806.5\tabularnewline
192 & 0 & $5_1+7_1,5_1+11_1$ & $A_2$ & 2603.0 & 248 & 1 & $4_1+6_1,10_1+6_1$ & $A_1$ & 2809.4\tabularnewline
193-194 & 1 & $6_1+8_1$ & $E$ & 2606.1 & 249-250 & 1 & {[}$4_1+8_1,10_1+6_1${]} & $E$ & 2810.3\tabularnewline
195-196 & 1,2 & $11_1+7_1$ & $E$ & 2616.1 & 251-252 & 1 & $4_1+8_1,10_1+8_1$ & $E$ & 2812.3\tabularnewline
197 & 0 & $10_1+11_1,4_1+7_1$ & $A_1$ & 2618.6 & 253 & 0 & {[}$5_1+11_1,5_1+7_1${]} & $A_1$ & 2816.2\tabularnewline
198-199 & 0 & $4_1+7_1,10_1+11_1$ & $E$ & 2621.5 & 254 & 2 & $11_1+7_1$ & $A_2$ & 2816.8\tabularnewline
200 & 0 & {[}$10_1+7_1,4_1+11_1${]} & $A_2$ & 2624.3 & 255 & 2 & $11_1+7_1$ & $A_1$ & 2817.8\tabularnewline
201-202 & 0 & $4_1+11_1,10_1+7_1$ & $E$ & 2629.9 & 256-257 & 0,2 & {[}$11_1+7_1,6_2${]} & $E$ & 2819.1\tabularnewline
203-204 & 0,2 & {[}$6_2,6_1+11_1${]} & $E$ & 2650.0 & 258-259 & 0 & {[}$10_1+11_1,10_1+7_1${]} & $E$ & 2836.3\tabularnewline
205-206 & 4 & $6_1$ & $E$ & 2667.3 & 260 & 2 & $5_1+8_1$ & $A_1$ & 2836.8\tabularnewline
207 & 0 & {[}$6_2,6_1+11_1${]} & $A_1$ & 2669.9 &  &  &  &  & 
\tabularnewline
\hline\hline
\end{tabular}
\footnotetext[1]{2566.163,2566.179}
\end{table}

\begin{table}
\caption{
Vibrational states of \ce{CH3OH}$\ldots$ [Table~\ref{tbl:VBO_I} continued]}\label{tbl:VBO_2}
\begin{tabular}{@{}ccccccrc@{}}
\hline
\hline
\# & $\nu_{\tau}$ & SAMs & $\Gamma$ & $\tilde{\nu}_\mr{PES}$ & $\delta$ & $\tilde{\nu}_\mr{exp}$ & \tabularnewline
\hline
261-262 & 5 & $8_1$ & $E$ & 2837.3 &  &  &  \tabularnewline
263-264 & 4 & $4_1,10_1$ & $E$ & 2841.7 &  &  &  \tabularnewline
265 & 0 & $3_1,5_2$ & $A_1$ & 2841.8 & [3.0] & \footnotemark[1]2844.7  & \citenum{Wang1998JCP_fit} \tabularnewline
266-267 & 4 & $5_1$ & $E$ & 2845.1 &  &  &  \tabularnewline
268-269 & 0 & {[}$3_1,5_2${]} & $E$ & 2850.7 & [3.1] & \footnotemark[1]2853.8  & \citenum{Wang1998JCP_fit} \tabularnewline
270-271 & 2 & $4_1+8_1$ & $E$ & 2862.0 &  &  &  \tabularnewline
272-273 & 2 & {[}$7_2,11_2${]} & $E$ & 2867.8 &  &  &  \tabularnewline
274-275 & 1 & $5_1+7_1,5_1+11_1$ & $E$ & 2870.7 &  &  &  \tabularnewline
276 & 4 & $10_1,4_1$ & $A_2$ & 2878.6 &  &  &  \tabularnewline
277 & 4 & $4_1,10_1$ & $A_1$ & 2879.6 &  &  &  \tabularnewline
278-279 & 1 & $4_1+11_1,10_1+7_1$ & $E$ & 2888.7 &  &  &  \tabularnewline
280 & 1 & {[}$4_1+7_1,10_1+11_1${]} & $A_2$ & 2891.9 &  &  &  \tabularnewline
281 & 1 & {[}$10_1+7_1,4_1+11_1${]} & $A_1$ & 2892.9 & [0.3] & \footnotemark[2]2893.2  & \citenum{Serrallach1974JMS} \tabularnewline
282-283 & 2 & {[}$6_1+7_1,6_1+11_1${]} & $E$ & 2900.4 &  &  &  \tabularnewline
284-285 & 0 & $4_1+5_1,10_1+5_1$ & $E$ & 2913.2 &   [--3] & \footnotemark[3]2910  & \citenum{Serrallach1974JMS}  \tabularnewline
286 & 0 & {[}$4_1+5_1,10_1+5_1${]} & $A_2$ & 2917.8 & [--6] & 2912  & \citenum{Serrallach1974JMS}  \tabularnewline
287 & 0 & $5_2$ & $A_1$ & 2920.9 &  & \tabularnewline
288-289 & 0,2 & {[}$10_1+5_1,5_1+11_1${]} & $E$ & 2921.5 &  &  &  \tabularnewline
290-291 & 2 & $5_1+7_1,5_1+11_1$ & $E$ & 2923.1 &  &  &  \tabularnewline
292 & 0 & {[}$4_1+5_1,4_1+10_1${]} & $A_1$ & 2928.4 & [1.1] & 2929.5  & \citenum{Serrallach1974JMS}  \tabularnewline
293-294 & 0 & {[}$10_1+5_1,2_1${]} & $E$ & 2930.3 &  &  &  \tabularnewline
295-296 & 0 & $5_2,4_2$ & $E$ & 2931.6 &  &  &  \tabularnewline
297 & 0 & $4_1+5_1,10_1+5_1$ & $A_2$ & 2933.9 &  &  &  \tabularnewline
298-299 & 2 & {[}$4_1+11_1,10_1+7_1${]} & $E$ & 2934.1 &  &  &  \tabularnewline
300 & 3 & $7_1+8_1,11_1+8_1$ & $A_1$ & 2935.1 &  &  &  \tabularnewline
301 & 3 & $7_1+8_1,11_1+8_1$ & $A_2$ & 2935.3 &  &  &  \tabularnewline
302-303 & 0 & {[}$4_1+10_1,4_2${]} & $E$ & 2938.5 &  &  &  \tabularnewline
304 & 0 & $4_1+10_1$ & $A_1$ & 2939.2 &  &  &  \tabularnewline
305 & 1 & $6_2$ & $A_2$ & 2940.3 &  &  &  \tabularnewline
306-307 & 1,2 & {[}$6_1+7_1,11_1+7_1${]} & $E$ & 2948.1 &  &  &  \tabularnewline
308 & 2 & $4_1+7_1,10_1+11_1$ & $A_1$ & 2952.2 &  &  &  \tabularnewline
309 & 2 & $10_1+7_1,4_1+11_1$ & $A_2$ & 2956.1 &  &  &  \tabularnewline
310 & 0 & {[}$4_2,10_2${]} & $A_1$ & 2957.0 & [1.4] & \footnotemark[1]2958.4  & \citenum{Wang1998JCP_fit} \tabularnewline
311-312 & 0 & {[}$4_1+10_1,2_1,9_1${]} & $E$ & 2959.1 & [2.1] & 2961.2  &  \citenum{Hanninen1999JCP_fit}\tabularnewline
313 & 0 & {[}$4_1+10_1,2_1,9_1${]} & $A_2$ & 2964.2 & [2.5] & 2966.7  & \citenum{Hanninen1999JCP_fit} \tabularnewline
314-315 & 0 & {[}$4_2,10_2${]} & $E$ & 2964.7 & [2.0] & \footnotemark[1]2966.7  & \citenum{Wang1998JCP_fit} \tabularnewline
316-317 & 5 & $7_1,11_1$ & $E$ & 2965.0 &  &  &  \tabularnewline
318 & 2 & $6_2$ & $A_1$ & 2966.6 &  &  &  \tabularnewline
319 & 5 & $7_1,11_1$ & $A_2$ & 2974.3 &  &  &  \tabularnewline
320 & 5 & $11_1,7_1$ & $A_1$ & 2974.3 &  &  &  \tabularnewline
321-322 & 2 & {[}$6_2,6_1+7_1${]} & $E$ & 2975.3 &  &  &  \tabularnewline
323-324 & 3 & {[}$7_1+8_1,11_1+8_1${]} & $E$ & 2989.0 &  &  &  \tabularnewline
325 & 1 & $5_1+6_1$ & $A_2$ & 2989.3 &  &  &  \tabularnewline
326-327 & 2 & $5_1+8_1$ & $E$ & 2994.7 &  &  &  \tabularnewline
328 & 2 & $5_1+6_1$ & $A_1$ & 2995.1 &  &  &  \tabularnewline
329-330 & 0 & {[}$2_1,9_1${]} & $E$ & 2999.8 & [3.9] & 3003.7  & \citenum{Xu1997JMS_CH_strech} \tabularnewline
331 & 0 & $9_1,2_1$ & $A_1$ & 3002.9 & [4.1] & 3007.0  & \citenum{Xu1997JMS_CH_strech} \tabularnewline
\hline
\hline
\end{tabular}
\footnotetext[1]{The values listed in Table I of Ref.~\citenum{Wang1998JCP_fit} are shifted by the zero-point energy of 128.1 \cm, taken from in Ref.~\citenum{Hanninen1999JCP_fit}.}
\footnotetext[2]{The assignment of Ref.~\citenum{Serrallach1974JMS} is $5_2$.}
\footnotetext[3]{The assignment of Ref.~\citenum{Serrallach1974JMS} is $10_2$.}
\end{table}

\begin{table}
\caption{
Vibrational states of \ce{CH3OH}$\ldots$ [Table~\ref{tbl:VBO_2} continued]}\label{tbl:VBO_3}
\begin{tabular}{@{}ccccc| ccccc@{}}
\hline
\hline
\# & $\nu_{\tau}$ & SAMs & $\Gamma$ & $\tilde{\nu}_\mr{PES}$ & \# & $\nu_{\tau}$ & SAMs & $\Gamma$ & $\tilde{\nu}_\mr{PES}$  \tabularnewline
\hline
332-333 & 1 & $4_1+6_1,10_1+6_1$ & $E$ & 3003.8 & 382 & 3,2 & {[}$11_1+7_1,6_1+11_1${]} & $A_1$ & 3157.4\tabularnewline
334-335 & 2 & $4_1+8_1,10_1+8_1$ & $E$ & 3006.6 & 383-384 & 1 & {[}$4_2,3_1${]} & $E$ & 3158.9\tabularnewline
336-337 & 2 & {[}$4_1+6_1,4_1+11_1${]} & $E$ & 3019.0 & 385 & 3 & $6_1+7_1,6_1+11_1$ & $A_2$ & 3159.3\tabularnewline
338-339 & 1 & {[}$5_1+6_1,5_1+6_1${]} & $E$ & 3021.2 & 386 & 3 & $6_1+11_1,6_1+7_1$ & $A_1$ & 3160.9\tabularnewline
340-341 & 3 & $6_1+8_1$ & $E$ & 3025.6 & 387 & 2 & {[}$5_1+11_1,5_1+7_1${]} & $A_2$ & 3162.4\tabularnewline
342 & 2 & $4_1+8_1,10_1+8_1$ & $A_2$ & 3028.2 & 388 & 0 & {[}$6_1+7_1+8_1,6_1+11_1+8_1${]} & $A_1$ & 3168.8\tabularnewline
343 & 2 & $10_1+8_1,4_1+8_1$ & $A_1$ & 3030.4 & 389-390 & 0 & {[}$7_1+8_2,6_1+7_1+8_1${]} & $E$ & 3170.8\tabularnewline
344 & 1 & {[}$10_1+6_1,4_1+6_1${]} & $A_2$ & 3034.3 & 391 & 1 & {[}$4_1+10_1,9_1${]} & $A_1$ & 3175.8\tabularnewline
345 & 1 & {[}$10_1+6_1,4_1+6_1${]} & $A_1$ & 3040.2 & 392-393 & 1,2 & {[}$4_1+11_1,4_1+10_1${]} & $E$ & 3176.7\tabularnewline
346-347 & 1 & {[}$4_1+6_1,10_1+6_1${]} & $E$ & 3042.1 & 394-395 & 2 & {[}$4_1+7_1,4_1+11_1${]} & $E$ & 3179.2\tabularnewline
348-349 & 1 & $3_1,5_2$ & $E$ & 3056.5 & 396 & 2 & {[}$5_1+6_1,5_1+11_1${]} & $A_1$ & 3187.6\tabularnewline
350-351 & 3 & $11_1+7_1$ & $E$ & 3071.5 & 397 & 1 & $2_1,9_1$ & $A_2$ & 3190.4\tabularnewline
352 & 0 & $8_3$ & $A_1$ & 3077.7 & 398-399 & 2 & $5_1+6_1$ & $E$ & 3191.4\tabularnewline
353-354 & 5 & $6_1$ & $E$ & 3084.7 & 400-401 & 2 & $6_2$ & $E$ & 3192.6\tabularnewline
355-356 & 0 & $8_3$ & $E$ & 3086.4 & 402-403 & 2 & $4_1+6_1,10_1+6_1$ & $E$ & 3195.1\tabularnewline
357-358 & 3 & $7_2,11_2$ & $E$ & 3102.0 & 404 & 2 & $3_1,5_2$ & $A_1$ & 3201.4\tabularnewline
359 & 3 & $8_2$ & $A_2$ & 3102.2 & 405-406 & 0 & $7_1+8_2,11_1+8_2$ & $E$ & 3202.2\tabularnewline
360 & 4 & $8_2$ & $A_1$ & 3103.2 & 407 & 0 & $7_1+8_2,11_1+8_2$ & $A_2$ & 3210.1\tabularnewline
361-362 & 2 & $5_1+7_1,5_1+11_1$ & $E$ & 3107.5 & 408-409 & 2 & {[}$4_1+6_1,4_1+7_1${]} & $E$ & 3210.4\tabularnewline
363 & 2 & {[}$10_1+11_1,4_1+7_1${]} & $A_2$ & 3109.0 & 410-411 & 1 & $4_1+5_1$ & $E$ & 3211.8\tabularnewline
364 & 2 & {[}$4_1+7_1,10_1+7_1${]} & $A_1$ & 3110.7 & 412 & 0 & {[}$6_1+7_2,6_1+11_2${]} & $A_1$ & 3214.1\tabularnewline
365 & 1 & $10_1+5_1,4_1+5_1$ & $A_2$ & 3121.2 & 413-414 & 0 & {[}$6_1+7_2,6_2+7_1${]} & $E$ & 3215.2\tabularnewline
366 & 0 & {[}$8_3,6_1+8_2${]} & $A_1$ & 3123.5 & 415 & 1 & $5_2$ & $A_2$ & 3215.9\tabularnewline
367-368 & 1 & {[}$4_2,4_1+10_1${]} & $E$ & 3124.0 & 416 & 2 & $4_1+6_1,10_1+6_1$ & $A_1$ & 3223.6\tabularnewline
369 & 1 & $10_1+5_1,4_1+5_1$ & $A_1$ & 3127.4 & 417-418 & 1 & {[}$4_1+10_1,4_2${]} & $E$ & 3223.6\tabularnewline
370-371 & 0 & {[}$8_3,7_1+8_2${]} & $E$ & 3127.9 & 419 & 2 & $4_1+6_1,10_1+6_1$ & $A_2$ & 3224.6\tabularnewline
372 & 1 & $3_1$ & $A_2$ & 3130.6 & 420-421 & 3 & $5_1+8_1$ & $E$ & 3225.5\tabularnewline
373-374 & 1 & {[}$4_1+5_1,10_1+5_1${]} & $E$ & 3133.4 & 422 & 5 & {[}$4_1,10_1${]} & $A_2$ & 3228.2\tabularnewline
375-376 & 1 & {[}$4_2,5_2${]} & $E$ & 3137.6 & 423 & 5 & {[}$4_1,10_1${]} & $A_1$ & 3228.3\tabularnewline
377 & 1 & {[}$4_1+10_1,10_1+5_1${]} & $A_1$ & 3141.3 & 424-425 & 3,4 & $7_1+8_1,11_1+8_1$ & $E$ & 3238.0\tabularnewline
378-379 & 2 & {[}$4_1+11_1,10_1+7_1${]} & $E$ & 3143.4 & 426-427 & 3 & $6_2$ & $E$ & 3238.7\tabularnewline
380 & 1,3 & {[}$4_1+10_1,11_1+7_1${]} & $A_2$ & 3146.7 & 428-429 & 5 & $5_1$ & $E$ & 3247.1\tabularnewline
381 & 3,2 & {[}$11_1+7_1,6_1+11_1${]} & $A_2$ & 3151.9 & 430-431 & 0 & {[}$11_1+7_1+8_1,6_1+7_1+8_1${]} & $E$ & 3247.5\tabularnewline
\hline
\hline
\end{tabular}
\end{table}

\begin{table}
\caption{
Vibrational states of \ce{CH3OH}$\ldots$ [Table~\ref{tbl:VBO_3} continued]}\label{tbl:VBO_4}
\makebox[\textwidth][c]{%
\begin{tabular}{@{}ccccc| ccccc@{}}
\hline
\hline
\# & $\nu_{\tau}$ & SAMs & $\Gamma$ & $\tilde{\nu}_\mr{PES}$ & \# & $\nu_{\tau}$ & SAMs & $\Gamma$ & $\tilde{\nu}_\mr{PES}$  \tabularnewline
\hline
432 & 1 & {[}$3_1,10_2${]} & $A_2$ & 3248.6 & 483 & 3 & $11_2,7_2$ & $A_2$ & 3391.0\tabularnewline
433 & 3 & {[}$4_1+8_1,10_1+8_1${]} & $A_1$ & 3252.2 & 484 & 4 & $7_2,11_2$ & $A_1$ & 3391.8\tabularnewline
434 & 3 & {[}$4_1+8_1,10_1+8_1${]} & $A_2$ & 3252.2 & 485-486 & 3 & {[}$10_1+11_1,4_1+7_1${]} & $E$ & 3393.8\tabularnewline
435 & 0,1 & {[}$7_1+8_1,11_1+7_1+8_1${]} & $A_2$ & 3256.7 & 487 & 1 & $8_3$ & $A_2$ & 3393.9\tabularnewline
436-437 & 3,4 & {[}$7_1+8_1,11_1+8_1${]} & $E$\footnotemark[1] & 3262.2 & 488-489 & 3 & {[}$5_1+7_1,5_1+11_1${]} & $E$ & 3397.9\tabularnewline
438-439 & 3 & {[}$4_1+8_1,10_1+8_1${]} & $E$\footnotemark[2] & 3262.5 & 490 & 0,1 & {[}$6_2+8_1,11_1+8_2${]} & $A_1$ & 3403.7\tabularnewline
440-441 & 1 & $2_1,9_1$ & $E$ & 3264.6 & 491-492 & 3 & {[}$4_1+11_1,10_1+7_1${]} & $E$\footnotemark[6] & 3405.6\tabularnewline
442-443 & 2 & $4_1+5_1,10_1+5_1$ & $E$\footnotemark[3] & 3273.3 & 493-494 & 0,1 & {[}$6_2+8_1,7_2+8_1${]} & $E$ & 3411.8\tabularnewline
444-445 & 7 & 0 & $E$\footnotemark[4] & 3273.8 & 495 & 2 & $4_1+10_1$ & $A_2$ & 3413.1\tabularnewline
446 & 5 & $8_1$ & $A_2$ & 3276.3 & 496 & 2 & $4_1+10_1$ & $A_1$ & 3415.3\tabularnewline
447 & 6 & $8_1$ & $A_1$ & 3276.3 & 497 & 0 & {[}$6_1+8_2,7_1+8_2${]} & $A_1$ & 3417.4\tabularnewline
448 & 2 & $5_2$ & $A_1$ & 3278.9 & 498-499 & 2 & $10_1+5_1,4_1+5_1$ & $E$ & 3417.9\tabularnewline
449-450 & 1 & $8_3$ & $E$ & 3286.0 & 500-501 & 5,6 & $7_1,11_1$ & $E$\footnotemark[7] & 3422.8\tabularnewline
451-452 & 2 & $4_1+10_1$ & $E$ & 3287.4 & 502-503 & 3,4 & $11_1+7_1$ & $E$\footnotemark[8] & 3425.2\tabularnewline
453-454 & 5 & $10_1,4_1$ & $E$\footnotemark[5] & 3290.9 & 504 & 3 & {[}$4_1+11_1,10_1+11_1${]} & $A_2$ & 3425.5\tabularnewline
455-456 & 0 & $6_1+11_1+7_1,11_2+7_1$ & $E$ & 3293.5 & 505 & 3 & {[}$10_1+11_1,4_1+11_1${]} & $A_1$ & 3425.9\tabularnewline
457 & 0 & {[}$11_2+7_1,6_1+11_1+7_1${]} & $A_2$ & 3301.9 & 506 & 0 & {[}$6_1+7_1+8_1,6_1+11_1+8_1${]} & $A_2$ & 3427.4\tabularnewline
458 & 2 & {[}$3_1,4_2${]} & $A_1$ & 3307.6 & 507-508 & 5,6 & {[}$7_1,11_1${]} & $E$ & 3428.6\tabularnewline
459 & 0 & {[}$11_1+7_1+8_1,11_2+8_1${]} & $A_1$ & 3323.4 & 509-510 & 1 & {[}$6_2+8_1,6_1+8_2${]} & $E$ & 3431.7\tabularnewline
460-461 & 2 & $2_1,9_1$ & $E$ & 3325.1 & 511-521 & 2 & {[}$5_2${]} & $E$ & 3433.5\tabularnewline
462-463 & 0 & {[}$7_2+8_1,11_2+8_1${]} & $E$ & 3328.9 & 513 & 2 & $8_3$ & $A_1$ & 3435.0\tabularnewline
464 & 3 & $6_1+8_1$ & $A_2$ & 3334.0 & 514-515 & 3 & $5_1+6_1$ & $E$ & 3438.5\tabularnewline
465 & 4 & $6_1+8_1$ & $A_1$ & 3334.2 & 516 & 2 & $4_1+5_1,10_1+5_1$ & $A_2$ & 3439.9\tabularnewline
466-467 & 0 & $6_1+8_2$ & $E$ & 3345.9 & 517 & 2 & $10_1+5_1,4_1+5_1$ & $A_1$ & 3441.6\tabularnewline
468-469 & 2 & {[}$3_1,5_2${]} & $E$ & 3352.6 & 518-519 & 4 & $8_2$ & $E$ & 3447.1\tabularnewline
470 & 3 & $5_1+7_1,5_1+11_1$ & $A_1$ & 3353.7 & 520-521 & 0 & {[}$11_2+7_1,7_3${]} & $E$ & 3449.5\tabularnewline
471 & 3 & $5_1+7_1,5_1+11_1$ & $A_2$ & 3353.7 & 522 & 0 & {[}$11_3,7_3${]} & $A_2$ & 3450.1\tabularnewline
472-473 & 2 & $3_1,5_2$ & $E$ & 3356.2 & 523 & 3 & $10_1+6_1,4_1+6_1$ & $A_2$ & 3452.3\tabularnewline
474 & 0,1 & {[}$6_1+8_2,11_1+8_2${]} & $A_1$ & 3358.1 & 524 & 3 & $10_1+6_1,4_1+6_1$ & $A_1$ & 3452.7\tabularnewline
475 & 1 & $8_3$ & $A_2$ & 3359.6 & 525 & 0 & {[}$6_3,6_2+11_1${]} & $A_1$ & 3454.4\tabularnewline
476 & 0 & $11_1+7_2$ & $A_1$ & 3367.0 & 526-527 & 3 & $5_1+6_1$ & $E$ & 3456.3\tabularnewline
477-478 & 3,4 & $11_1+7_1$ & $E$ & 3375.3 & 528-529 & 2 & $3_1,10_2$ & $E$ & 3462.9\tabularnewline
479-480 & 3,4 & {[}$11_1+7_1,7_2${]} & $E$ & 3381.7 & 530-531 & 0 & {[}$7_3,6_3${]} & $E$ & 3463.5\tabularnewline
481-482 & 1 & $6_1+8_2$ & $E$ & 3386.4 &  &  &  &  & \tabularnewline
\hline
\hline
\end{tabular}
}
\footnotetext[1]{3262.157, 3262.172\vspace{0.02cm}}
\footnotetext[2]{3262.459, 3262.476\vspace{0.02cm}}
\footnotetext[3]{3273.340, 3273.368\vspace{0.02cm}}
\footnotetext[4]{3273.796, 3273.809\vspace{0.02cm}}
\footnotetext[5]{3290.923, 3290.960\vspace{0.02cm}}
\footnotetext[6]{3405.636, 3405.649\vspace{0.02cm}}
\footnotetext[7]{3422.825, 3422.884\vspace{0.02cm}}
\footnotetext[8]{3425.241, 3425.260\vspace{0.02cm}}
\end{table}

\begin{table}
\caption{
Vibrational states of \ce{CH3OH}$\ldots$ [Table~\ref{tbl:VBO_4} continued]}\label{tbl:VBO_5}
\makebox[\textwidth][c]{%
\begin{tabular}{@{}ccccc| ccccc@{}}
\hline
\hline
\# & $\nu_{\tau}$ & SAMs & $\Gamma$ & $\tilde{\nu}_\mr{PES}$ & \# & $\nu_{\tau}$ & SAMs & $\Gamma$ & $\tilde{\nu}_\mr{PES}$  \tabularnewline
\hline
532-533 & 1 & {[}$7_1+8_2,11_1+7_1+8_1${]} & $E$ & 3471.8 & 581-582 & 0 & {[}$4_1+11_1+8_1,4_1+8_2${]} & $E$ & 3570.4\tabularnewline
534 & 2 & $2_1,9_1$ & $A_2$ & 3472.3 & 583-584 & 1 & {[}$7_2+8_1,11_2+8_1${]} & $E$ & 3571.9\tabularnewline
535-536 & 3,4 & $6_1+11_1,6_1+7_1$ & $E$ & 3474.3 & 585 & 0 & {[}$10_1+8_2,4_1+8_2${]} & $A_2$ & 3573.4\tabularnewline
537 & 2 & $2_1,9_1$ & $A_1$ & 3476.8 & 586-587 & 1 & {[}$11_2+7_1,7_3${]} & $E$ & 3575.2\tabularnewline
538 & 2 & {[}$2_1,9_1${]} & $A_1$ & 3478.3 & 588 & 0,2 & {[}$6_1+8_2,6_2+8_1${]} & $A_1$ & 3577.9\tabularnewline
539-540 & 0,1 & {[}$6_1+7_1+8_1,11_1+8_2${]} & $E$ & 3480.5 & 589-590 & 0,1 & {[}$4_1+7_1+8_1,11_2+7_1${]} & $E$ & 3578.9\tabularnewline
541-542 & 3 & $10_1+6_1,4_1+6_1$ & $E$\footnotemark[1] & 3481.3 & 591 & 0 & {[}$10_1+8_2,4_1+8_2${]} & $A_1$ & 3580.6\tabularnewline
543-544 & 3 & {[}$6_2,6_1+11_1${]} & $E$ & 3490.0 & 592 & 0 & {[}$5_1+6_1+7_1,5_1+6_1+11_1${]} & $A_1$ & 3586.3\tabularnewline
545-546 & 2 & {[}$4_1+10_1,2_1${]} & $E$ & 3492.4 & 593-594 & 0 & {[}$5_1+6_1+11_1,5_1+7_2${]} & $E$ & 3587.6\tabularnewline
547-548 & 3,4 & {[}$6_2,6_1+7_1${]} & $E$ & 3493.2 & 595 & 1 & $6_1+8_2$ & $A_2$ & 3590.7\tabularnewline
549 & 1 & {[}$11_3,7_3${]} & $A_2$ & 3495.0 & 596-597 & 4 & $7_1+8_1,11_1+8_1$ & $E$\footnotemark[2] & 3592.2\tabularnewline
550 & 0 & $5_1+8_2$ & $A_1$ & 3499.9 & 598-599 & 3 & $3_1,5_2$ & $E$ & 3594.2\tabularnewline
551 & 1 & {[}$6_1+7_1+8_1,6_1+11_1+8_1${]} & $A_2$ & 3502.7 & 600 & 4 & {[}$7_1+8_1,11_1+8_1${]} & $A_2$ & 3594.4\tabularnewline
552-553 & 0 & {[}$5_1+8_2,5_1+8_2${]} & $E$ & 3508.2 & 601-602 & 2 & $8_3$ & $E$ & 3596.3\tabularnewline
554-555 & 0,2 & {[}$11_1+7_1+8_1,6_1+7_1+8_1${]} & $E$ & 3517.3 & 603 & 4 & $11_1+8_1,7_1+8_1$ & $A_1$ & 3601.4\tabularnewline
556-557 & 0,2 & {[}$7_1+8_2,11_1+8_2${]} & $E$ & 3522.4 & 604-605 & 1 & {[}$11_1+7_2,11_3${]} & $E$ & 3603.2\tabularnewline
558 & 3 & $5_1+8_1$ & $A_2$ & 3526.3 & 606 & 4 & {[}$11_1+8_1,7_1+8_1${]} & $A_1$ & 3605.7\tabularnewline
559 & 4 & $5_1+8_1$ & $A_1$ & 3527.1 & 607-608 & 1,2 & {[}$7_1+8_2,11_1+7_1+8_1${]} & $E$ & 3608.0\tabularnewline
560-561 & 0 & $4_1+8_2,10_1+8_2$ & $E$ & 3527.9 & 609-610 & 0 & {[}$10_1+6_1+7_1,10_1+7_2${]} & $E$ & 3610.4\tabularnewline
562-563 & 1,2 & {[}$7_1+8_2,11_1+8_2${]} & $E$ & 3530.2 & 611 & 4 & $7_1+8_1,11_1+8_1$ & $A_2$ & 3611.4\tabularnewline
564-565 & 3,4 & $4_1+8_1,10_1+8_1$ & $E$ & 3533.4 & 612 & 0 & {[}$4_1+6_1+11_1,10_1+6_1+7_1${]} & $A_2$ & 3612.1\tabularnewline
566 & 0 & $10_1+8_2,4_1+8_2$ & $A_2$ & 3535.5 & 613 & 2 & $6_1+8_2$ & $A_1$ & 3616.2\tabularnewline
567 & 1 & {[}$6_2+11_1,6_2+7_1${]} & $A_1$ & 3536.2 & 614-615 & 0 & {[}$10_1+6_1+11_1,10_1+11_2${]} & $E$ & 3619.0\tabularnewline
568-569 & 0 & $4_1+8_2,10_1+8_2$ & $E$ & 3537.9 & 616 & 0 & {[}$10_1+6_1+11_1,4_1+6_1+7_1${]} & $A_1$ & 3621.0\tabularnewline
570 & 0 & $10_1+8_2,4_1+8_2$ & $A_1$ & 3541.0 & 617-618 & 0 & {[}$5_1+11_1+8_1,5_1+7_1+8_1${]} & $E$ & 3621.2\tabularnewline
571 & 0 & {[}$5_1+6_1+8_1,5_1+7_1+8_1${]} & $A_1$ & 3544.5 & 619 & 0 & $5_1+11_1+8_1,5_1+7_1+8_1$ & $A_2$ & 3627.4\tabularnewline
572 & 0,1 & {[}$11_2+7_1,6_2+7_1${]} & $A_2$ & 3545.3 & 620-621 & 1 & $6_1+8_2$ & $E$ & 3628.2\tabularnewline
573 & 5 & $6_1$ & $A_2$ & 3547.6 & 622-623 & 0,2 & {[}$11_1+7_1+8_1,11_3${]} & $E$ & 3630.0\tabularnewline
574 & 6 & $6_1$ & $A_1$ & 3547.7 & 624-625 & 1,2 & {[}$7_2+8_1,11_1+7_1+8_1${]} & $E$ & 3639.5\tabularnewline
575-576 & 0 & {[}$5_1+7_1+8_1,5_1+8_2${]} & $E$ & 3548.1 & 626 & 0 & $4_1+7_1+8_1,10_1+11_1+8_1$ & $A_1$ & 3646.3\tabularnewline
577 & 4 & $6_2$ & $A_1$ & 3561.9 & 627-628 & 0 & {[}$4_1+7_1+8_1,10_1+11_1+8_1${]} & $E$ & 3649.1\tabularnewline
578 & 3 & $6_2$ & $A_2$ & 3562.5 & 629 & 0 & {[}$4_1+11_1+8_1,10_1+7_1+8_1${]} & $E$ & 3650.7\tabularnewline
579-580 & 3,4 & $4_1+8_1,10_1+8_1$ & $E$ & 3565.4 & 630-631 & 3 & {[}$4_1+11_1,10_1+7_1${]} & $E$ & 3650.9\tabularnewline
\hline
\hline
\end{tabular}
}
\footnotetext[1]{3481.281, 3481.294\vspace{0.02cm}}
\footnotetext[2]{3592.206, 3592.218\vspace{0.02cm}}
\end{table}

\begin{table}
\caption{
Vibrational states of \ce{CH3OH}$\ldots$ [Table~\ref{tbl:VBO_5} continued]}\label{tbl:VBO_6}
\begin{tabular}{@{}cccccccc@{}}
\hline
\hline
\# & $\nu_{\tau}$ & SAMs & $\Gamma$ & $\tilde{\nu}_\mr{PES}$ & $\delta$ & $\tilde{\nu}_\mr{exp}$ & \tabularnewline
\hline
632-633 & 3 & $4_1+10_1$ & $E$ & 3651.4 &  &  & \tabularnewline
634-635 & 3,4 & {[}$5_1+7_1,5_1+11_1${]} & $E$ & 3653.8 &  &  & \tabularnewline
636 & 3 & $10_1+5_1,4_1+5_1$ & $A_2$ & 3654.2 &  &  & \tabularnewline
637 & 3 & $10_1+5_1,4_1+5_1$ & $A_1$ & 3654.6 &  &  & \tabularnewline
638-639 & 0 & $4_1+11_1+8_1,10_1+7_1+8_1$ & $E$ & 3657.5 &  &  & \tabularnewline
640-641 & 0,2 & {[}$11_1+7_1+8_1,6_3${]} & $E$ & 3660.9 &  &  & \tabularnewline
642-643 & 0 & {[}$5_1+11_1+7_1,5_1+11_2${]} & $E$ & 3665.4 &  &  & \tabularnewline
644 & 1 & {[}$11_1+7_2,11_3${]} & $A_2$ & 3665.6 &  &  & \tabularnewline
645-646 & 3 & $5_2$ & $E$ & 3672.9 &  &  & \tabularnewline
647 & 0 & $6_3$ & $A_1$ & 3673.2 &  &  & \tabularnewline
648 & 0,1 & {[}$5_1+11_1+7_1,5_1+11_1+7_1${]} & $A_2$ & 3673.7 &  &  & \tabularnewline
649-650 & 0 & {[}$6_2+8_1,6_2+8_1${]} & $E$ & 3678.5 &  &  & \tabularnewline
651-652 & 3,4 & $5_1+7_1,5_1+11_1$ & $E$ & 3682.2 &  &  & \tabularnewline
653-654 & 5,6 & {[}$4_1,10_1${]} & $E$ & 3682.3 &  &  & \tabularnewline
655-656 & 3 & {[}$4_1+5_1,10_1+5_1${]} & $E$ & 3682.7 &  &  & \tabularnewline
657 & 3 & {[}$10_1+7_1,4_1+11_1${]} & $A_1$ & 3684.8 &  &  & \tabularnewline
658 & 3 & $4_1+7_1,10_1+11_1$ & $A_2$ & 3685.0 &  &  & \tabularnewline
659 & 4 & $10_1+11_1,4_1+7_1$ & $A_1$ & 3685.3 &  &  & \tabularnewline
660 & 0 & $1_1$ & $A_1$ & 3686.4 & [--1.1] & 3685.3 & \citenum{Hunt1998JMS_OH_stretch} \tabularnewline
661 & 0,4 & {[}$10_1+11_1+7_1,10_1+7_1${]} & $A_2$ & 3687.5 &  &  & \tabularnewline
662-663 & 0,2 & {[}$7_3,11_3${]} & $E$ & 3688.2 &  &  & \tabularnewline
664 & 3 & {[}$2_1,9_1${]} & $A_2$ & 3688.6 &  &  & \tabularnewline
665 & 3 & {[}$2_1,9_1${]} & $A_1$ & 3688.7 &  &  & \tabularnewline
666-667 & 0 & {[}$4_1+11_1+7_1,4_1+6_1+7_1${]} & $E$ & 3689.9 &  &  & \tabularnewline
668-669 & 0 & $1_1$ & $E$ & 3692.6 & [--1.0] & 3691.6 & \citenum{Hunt1998JMS_OH_stretch} \tabularnewline
670 & 5 & $5_1$ & $A_2$ & 3692.7 &  &  & \tabularnewline
671 & 6 & $5_1$ & $A_1$ & 3692.8 &  &  & \tabularnewline
672-673 & 4 & $6_1+8_1$ & $E$ & 3695.1 &  &  & \tabularnewline
674 & 4 & {[}$4_1+7_1,10_1+11_1${]} & $E$ & 3696.4 &  &  & \tabularnewline
675 & 4 & {[}$10_1+7_1,4_1+11_1${]} & $A_2$ & 3696.7 &  &  & \tabularnewline
\hline
\hline
\end{tabular}
\end{table}